\pdfoutput=1
\documentclass[a4paper,fleqn,usenatbib]{mnras}

\usepackage{newtxtext,newtxmath}

\usepackage[T1]{fontenc}
\usepackage{ae,aecompl}

\usepackage{multirow}
\usepackage{pgfplots}
\usepackage{subfig}
\usepackage{mathtools, cuted}

\title[An optimal nonlinear neutrino method]{An optimal nonlinear method for simulating relic neutrinos}

\author[Elbers, Frenk, Jenkins, Li, and Pascoli]{Willem Elbers$^{1}$, Carlos S. Frenk$^{1}$, Adrian Jenkins$^{1}$, Baojiu Li$^{1}$, and Silvia Pascoli$^{2}$\\
$^{1}$Institute for Computational Cosmology, Department of Physics, Durham University, South Road, Durham, DH1 3LE, UK\\
$^{2}$Institute for Particle Physics Phenomenology, Department of Physics, Durham University, South Road, Durham, DH1 3LE, UK}

\date{Last updated 14 October, 2020; in original form 14 October, 2020}

\pubyear{2020}

\pubyear{2020}

\definecolor{lightgrey}{RGB}{230, 230, 230}

\begin{document}
\label{firstpage}
\pagerange{\pageref{firstpage}--\pageref{lastpage}}
\maketitle

\begin{abstract}
Cosmology places the strongest current limits on the sum of neutrino masses. Future observations will further improve the sensitivity and this will require accurate cosmological simulations to quantify possible systematic uncertainties and to make predictions for nonlinear scales, where much information resides. However, shot noise arising from neutrino thermal motions limits the accuracy of simulations. In this paper, we introduce a new method for simulating large-scale structure formation with neutrinos that accurately resolves the neutrinos down to small scales and significantly reduces the shot noise. The method works by tracking perturbations to the neutrino phase-space distribution with particles and reduces shot noise in the power spectrum by a factor of $\mathcal{O}\left(10^2\right)$ at $z=0$ for minimal neutrino masses and significantly more at higher redshifts, without neglecting the back-reaction caused by neutrino clustering. We prove that the method is part of a family of optimal methods that minimize shot noise subject to a maximum deviation from the nonlinear solution. Compared to other methods we find permille level agreement in the matter power spectrum and percent level agreement in the large-scale neutrino bias, but large differences in the neutrino component on small scales. A basic version of the method can easily be implemented in existing $N$-body codes and makes it possible to run neutrino simulations with significantly reduced particle load. Further gains are possible by constructing background models based on perturbation theory. A major advantage of this technique is that it works well for all masses, enabling a consistent exploration of the full neutrino parameter space.
\end{abstract}




\section{Introduction}

The discovery of neutrino masses \citep{fukuda98,ahmad02,eguchi03} calls for extensions of the Standard Model of particle physics and provides the only known form of dark matter. Measuring the masses is crucial for understanding their origin and for constraining cosmological parameters. While the neutrino mass squared differences are known to a few percent, the absolute masses are unknown and there remain two possible mass orderings: normal and inverted. A rich experimental programme is aimed at determining the mass ordering, measuring the mass scale set by the lightest neutrino and completing the overall picture of neutrino properties. Cosmology plays a vital role in this programme due its ability to provide an independent and complementary constraint on the sum of neutrino masses, $\sum m_\nu$ \citep{bond80,hu98} with a potential sensitivity below 0.02 eV \citep{font14,chudaykin19,sprenger19}.

Ongoing and planned neutrino experiments will establish the mass ordering with a discovery expected by the end of the decade. Although oscillation data have shown persistent hints of normal ordering, this preference has decreased to $1.6\sigma$ over the past year \citep{esteban20}. The mass ordering can be established by exploiting matter effects in long baseline neutrino oscillation experiments, as in \textsc{dune} \citep{acciarri15}, and in the Earth for atmospheric neutrinos, as in \textsc{orca} \citep{adrian16} and \textsc{Hyper-K} \citep{abe11}, as well as vacuum oscillations in medium baseline reactor neutrino experiments, specifically \textsc{Juno} \citep{an16}. Each approach is challenging, so information from multiple sources is essential. Single $\beta$-decay is the experiment of choice for direct mass searches and provides a model-independent determination of their value. The \textsc{katrin} experiment is ongoing and has put a bound of $m_\beta<1.1\text{ eV}$ on the value of quasi-degenerate neutrino masses with the aim of reaching $m_\beta<0.2$ eV in the near future \citep{aker19}. Project 8 will have the potential to set a limit of $m_\beta<0.04$ eV \citep{esfahani17}. Neutrinoless double $\beta$-decay can also provide information on neutrino masses \citep{bilenky01,pascoli02,nunokawa02}, albeit entangled with the value of the Majorana CP-violating phases and affected by uncertainty in the nuclear matrix elements \citep{vergados16}. For a recent review see e.g. \cite{giuliani19}.

The complementarity between these different strategies is of great interest. A cosmological measurement of  $\sum m_\nu$ would provide a target for direct mass searches \citep{drexlin13,mertens16}. An incompatibility between the two would indicate a non-standard cosmological evolution or new neutrino properties. A cosmological bound of $\sum m_\nu<100$ meV would suggest a normal mass ordering, which should be confronted with evidence from neutrino experiments. Finally, there is a strong synergy with neutrinoless double $\beta$-decay. Knowing the mass ordering and the sum of neutrino masses would narrow down the range of values for the effective Majorana mass parameter, providing a clear target for future experiments. 

Measuring the mass scale, and potentially ruling out the inverted mass ordering, is therefore a major target of near-term cosmological surveys, including \textsc{desi} \citep{desi13}, \textsc{euclid} \citep{euclid11}, and \textsc{lsst} \citep{lsst09}. In order to analyse these surveys and to extract a mass measurement, there has been a substantial effort to model precisely the effects of massive neutrinos on structure formation. From the analytical side, a swathe of new techniques such as time RG perturbation theory \citep{upadhye19} and effective field theories \citep{senatore17,colas19}, promise to push the validity of perturbation theory into the quasi-linear r\'egime. In the nonlinear r\'egime, $N$-body simulations offer the most accurate picture of structure formation. Yet incorporating neutrinos into $N$-body simulations has proved to be a challenge and some doubts remain about the validity of neutrino simulations on small scales.

The main obstacle to simulating neutrinos is that, in contrast to cold dark matter and baryons, neutrinos have a significant velocity dispersion. This effectively turns the 3-dimensional problem of structure formation, for which $N$-body simulations are well suited, into a 6-dimensional phase-space problem. If no provisions are made, a far greater number of simulation particles is needed to sample properly the phase-space manifold. A further complication arises from the fact that neutrinos are relativistic at high redshifts, such that simulations need to handle both the r\'egime where neutrinos are best described as radiation and the r\'egime where neutrinos are better described as massive particles.

The first neutrino simulations were carried out by \citet{klypin83} and \citet{frenk83}, when neutrinos were thought to be much more massive and the velocity dispersion not as problematic. Modern simulations with sub-electronvolt neutrinos were pioneered by \citet{brandbyge08,viel10}. Neutrinos are most commonly included in simulations as particles whose initial velocity is the sum of a peculiar gravitational component and a random component sampled from a Fermi-Dirac distribution \citep{brandbyge08,viel10,bird12,villaescusa12,adamek17,banerjee18,emberson17,inman15,villaescusa15,castorina15,villaescusa20}. The main difficulty with particle simulations is shot noise caused by the velocity dispersion. This problem is more severe for the smallest neutrino masses, which are observationally most relevant. Because neutrinos are a subdominant component, the error in the total matter distribution is relatively small. However, shot noise obscures the small-scale behaviour of the neutrinos and is clearly undesirable if one is interested in the neutrino component and its effect on structure formation.

To overcome the problems with particle simulations, grid simulations evolve the neutrino distribution using a system of fluid equations, which requires a scheme to close the moment hierarchy at some low order \citep{brandbyge09,viel10,hannestad12,archidiacono16,banerjee16,dakin19,tram19,inman20}, or as a linear response to the non-relativistic matter density \citep{ali12,liu18,mccarthy18}. Even more efficiently, but in the same spirit of treating neutrinos perturbatively, the total effect of neutrinos has been included as a post-processing step in the form of a gauge transformation \citep{partmann20}. While these approaches do not suffer from shot noise, they are not able to capture the full nonlinear evolution of the neutrinos at late times. This problem becomes more severe for more massive neutrinos, but is present even for minimal neutrino masses. A number of hybrid simulations have therefore combined grid and particle methods \citep{brandbyge10,banerjee16,bird18}, typically transitioning from a fluid method to a particle method at some redshift when the neutrinos become nonlinear. Another interesting alternative is to integrate the Poisson-Boltzmann equations directly on the grid \citep{yoshikawa20}.

The method proposed in this paper can be considered as a type of hybrid method that integrates neutrino particles but only uses the information contained in the particles to the extent that it is necessary. This is accomplished by dynamically transitioning from a smooth background model to a nonlinear model at the individual particle level. It relies on the noiseless (but approximate) background model as much as possible, thereby producing the smallest amount of shot noise possible whilst solving the full nonlinear system. The main idea is to decompose the phase-space distribution function $f(x,p,t)$ into a background model $f_0(x,p,t)$ which can be solved without noise, and a perturbation which is carried by the simulation particles:
\begin{align*}
	f(x,p,t) = f_0(x,p,t) + \delta f(x,p,t).
\end{align*}

\noindent
The choice of background model is arbitrary, but the method performs best whenever $f_0(x,p,t)$ is strongly correlated with $f(x,p,t)$, in a way that will be made precise below. If the choice of background model is poor, the method performs no worse than an ordinary $N$-body simulation, except for the small amount of overhead associated with evaluating $f_0(x,p,t)$. Note that the background model is just an approximation of $f$ and can itself be a perturbed Fermi-Dirac distribution.

This type of method has a long history in other fields and is variably known as the method of `perturbation particles' or more commonly as the `$\delta f$ method', which is the name we shall adopt. \citet{merritt87} and \citet{leeuwin93} discussed the method of perturbation particles in stellar dynamics. Around the same time, the $\delta f$ method arose in plasma physics \citep{tajima83,parker93,dimits93,aydemir94}. While the method of perturbation particles is not widely known today in astronomy, the $\delta f$ method is standard fare in plasma physics. A major difficulty in astronomical applications is the absence of a background model that captures enough of the dynamics to be useful. In contrast, plasma physicists are often interested in turbulent phenomena arising in an otherwise stable system, with a natural candidate for a background model $f_0$ at hand. Our work is motivated by the fact that there is also a natural background model for cosmic neutrinos, namely the phase-space density predicted by perturbation theory. There is a major synergy between $\delta f$ $N$-body simulations proposed here and work on improved perturbation theory methods. A better background model means a smaller dependence on the particles and therefore further reduced shot noise. We will show however that even the 0$^\text{th}$ order approximation, which is just a homogeneous redshifted Fermi-Dirac distribution, provides a significant improvement over ordinary $N$-body methods.

The remainder of the paper is structured as follows. In Section~\ref{sec:mc}, we derive the $\delta f$ method and describe its use as a variance reduction method for $N$-body simulations. We also show that the method is part of a family of optimal hybrid methods. In Section~\ref{sec:verification}, we illustrate the method with a one-dimensional test problem. In Section~\ref{sec:relativity}, we discuss how the method can be embedded in relativistic simulations. Our suite of simulations is then described in Section~\ref{sec:sims}. The method is compared with commonly used alternatives in Section~\ref{sec:results}. We consider higher order background models based on perturbation theory in Section~\ref{sec:higher}. Finally, we conclude in Section~\ref{sec:fin}.

\section{Derivation}\label{sec:mc}

The phase-space evolution of self-gravitating collisionless particles is described by the Poisson-Boltzmann equations, which in the single-fluid case read
\begin{align*}
	Lf &\equiv \left[\frac{\partial}{\partial t} + p\cdot\nabla - \nabla\Phi \cdot \nabla_p\right] f = 0,\\
	\nabla^2\Phi = 4\pi G\rho &= 4\pi G\int\mathrm{d}^3p\sqrt{m^2+p^2} f(x,p,t).
\end{align*}

\noindent
Here, $\Phi$ is the gravitational potential, $\rho$ the energy density, and $f$ the phase-space density. In general, the Liouville operator $L$ acts on each fluid separately and the potential should be summed over all fluid components. In relativistic perturbation theory, this system can be written as a hierarchy of moment equations for the neutrinos, which is solved to first order with Boltzmann codes such as \textsc{class} \citep{lesgourgues11} or \textsc{camb} \citep{lewis11}. To extend our predictions to the nonlinear r\'egime, we can use $N$-body codes, which solve the Poisson-Boltzmann system by the method of characteristics. Characteristic curves satisfy
\begin{align*}
	\frac{\mathrm{d}x}{\mathrm{d}t} = p \;\,\;\,\;\,\;\,\text{and}\;\,\;\,\;\,\;\,\frac{\mathrm{d}p}{\mathrm{d}t} = -\nabla\Phi.
\end{align*}

\noindent
By construction, one finds that $\mathrm{d}f/\mathrm{d}t=Lf=0$ along these curves. To infer statistics of the phase-space distribution, such as the number density, $n(x,t)=\langle1\rangle_p$, we simulate $N$ of these trajectories. Let $g(x,p)$ be the initial sampling distribution of Lagrangian marker particles in phase space. Then, a phase-space statistic is given by
\begin{align}
	A(x,t) = \langle A\rangle_p &= \int\mathrm{d}^3p\;f(x,p,t)A(x,p,t) \nonumber\\
	 	&\cong \frac{1}{N}\sum_{i=1}^N \frac{f(x_i,p_i,t)}{g(x_i,p_i)}A(x_i,p_i,t).\label{eq:estimator}
\end{align}

\noindent
The usual approach is to set $g(x,p) = f(x,p,t_0)$, in which case the sum reduces to a simple average over marker particles. The error in our estimate of $A$ is then $\sigma_A/\sqrt{N}$. Hence, if the
distribution, $f(x,p,t)$, has a large intrinsic scatter, we need a large $N$ to beat down the noise. Alternatively, we might construct an estimator with a smaller error. Let us therefore write the phase-space distribution function, $f$, as a background model, $f_0$, together with some perturbation, $\delta f$:
\begin{align*}
	f(x,p,t) = f_0(x,p,t) + \delta f(x,p,t).
\end{align*}

\noindent
We can reduce the error by only using the particles to estimate the
perturbed distribution, $\delta f$. We replace \eqref{eq:estimator}
with
\begin{align*}
	A(x,t) &= \int\mathrm{d}^3p\;\big[f_0(x,p,t)+\delta f(x,p,t)\big]A(x,p,t)\\
	 	&\cong A_0(x,t) + \frac{1}{N}\sum_{i=1}^N \frac{\delta f(x_i,p_i,t)}{g(x_i,p_i)}A(x_i,p_i,t).
\end{align*}

\noindent
This is useful if
\begin{align*}
A_0(x,t)=\int\mathrm{d}^3p\; f_0(x,p,t) A(x,p,t)
\end{align*}

\noindent
can be computed efficiently and if $f$ and $f_0$ are strongly correlated, so that the second term is small. The simplest choice of background model is a homogeneous Fermi-Dirac distribution
\begin{align}
	f_0(x,p,t) = \frac{g_s}{(2\pi)^3}\frac{1}{e^{ap/(k_bT)}+1}, \label{eq:homo_fd}
\end{align}

\noindent
with $g_s$ internal degrees of freedom. Here, $a=a(t)$ is the scale factor and $ap$ the present-day momentum. Since the noise reduction scales with the correlation between $f_0$ and $f$, we can achieve further gains by adding more information to the background model. The obvious next step is to use perturbation theory to improve on \eqref{eq:homo_fd}. This option is considered in section \ref{sec:higher}.

\subsection{Implementation}\label{sec:implement}

To implement the $\delta f$ method in cosmological $N$-body simulations, we simply replace the particle mass with a weighted mass:
\begin{align*}
	m \to m w_i = m\frac{\delta f(x_i,p_i,t)}{g(x_i,p_i)}.
\end{align*}

\noindent
The weights are computed by comparing the true phase-space density with the background model. We know the true density for each particle, because $Lf=Lg=0$ along characteristic curves. We record the two numbers $f$ and $g$ at the initial position of each particle and use these to compute the weights, $w_i$, during each subsequent step. We note that any sampling distribution $g(x,p)$ is valid provided that $g\neq 0$ almost everywhere $f\neq 0$. We will continue to use the common choice, $g=f$, where $f$ is the (perturbed) Fermi-Dirac distribution. In general, the optimal choice of $g$ will depend on the phase-space statistic of interest. Choosing a distribution, $g$, that oversamples slower particles can lead to an additional reduction in shot noise.

Given the homogeneous Fermi-Dirac background model \eqref{eq:homo_fd}, the neutrino density becomes
\begin{align*}
	\rho_\nu(x,t) &= \bar{\rho}_\nu(t) + \sum_{i=1}^N mw_i\,\delta^{(3)}(x-x_i).
\end{align*}

\noindent
Cosmological $N$-body simulations only compute the perturbed potential, since the background density $\bar{\rho}$ is accounted for in the background equations. The only change affecting the force calculation is therefore the weighting of the particles.

\begin{figure}
	\normalsize
	\centering
	\includegraphics{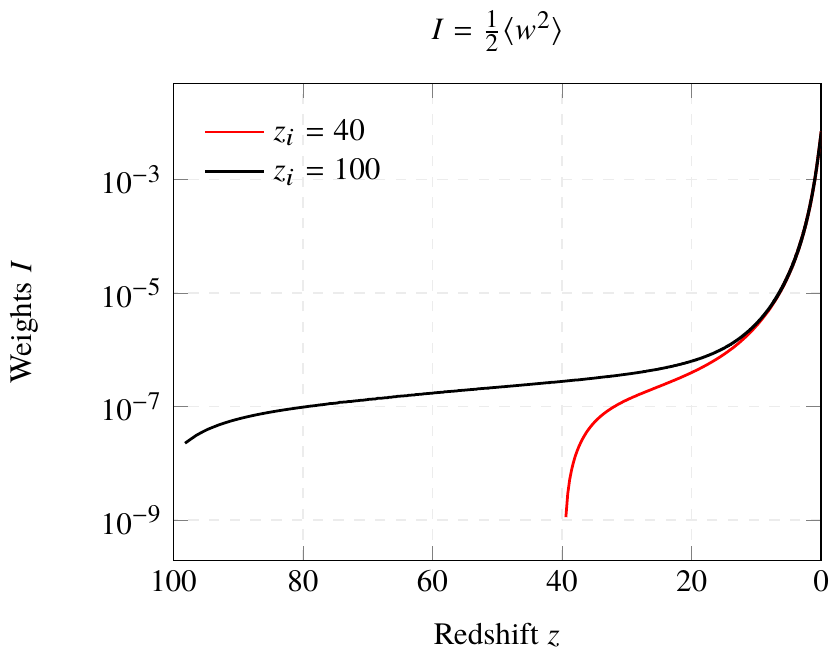}
	\caption{Evolution of particle weights for a $\sum m_\nu=100$ meV cosmology, starting at different redshifts $z_i$. The mean squared particle weight $\langle w^2\rangle$ represents the effective reduction in shot noise.}
	\label{fig:weights_time}
\end{figure}

The mean squared weight, $I=\tfrac{1}{2}\langle w^2\rangle$, is a convenient statistic to quantify the importance of including the neutrino particles. We show the evolution of $I$ for a $\sum m_\nu=100$ meV simulation with the homogeneous background model \eqref{eq:homo_fd} in Fig.~\ref{fig:weights_time}. At early times, particles deviate very little from their initial trajectory and the weights are negligible. We find that $I=4\times 10^{-7}$ at $z=20$,
$I=3\times10^{-6}$ at $z=10$, and $I=2\times10^{-5}$ at $z=5$. This early reduction is important as shot noise at high redshifts inhibits the growth of physical structure and can seed additional fluctuations that grow by gravitational instability. At late times, when nonlinear effects become important, the weights increase to $I=2\times10^{-4}$ at $z=2$, $I=1\times 10^{-3}$ at $z=1$, and $I=6.7\times10^{-3}$ at $z=0$, independently of the starting redshift of the simulation. This translates to a reduction in shot noise, $\sigma = 2VI/N$, or an
effective increase in particle number at $z=0$ by a factor $(2I)^{-1}=75$. Finally, we note that one can save computational resources by integrating only a fraction of the neutrino particles as long as $I$ remains small. We do not consider this possibility here.

\subsection{Variance reduction}\label{sec:var_reduct}

The $\delta f$ method is an application of the much more general control variates method \citep{ross12,aydemir94}. This is a variance reduction technique commonly used in Monte Carlo simulations. See \citet{chartier20} for another recent application in cosmology. We briefly review the method here. Let $A$ be a random variable with an unknown expectation $\mathrm{E}[A]=\mathcal{A}$. Given independent random samples $A_i$, the standard estimator is given by
\begin{align*}
	\hat{\mathcal{A}} = \frac{1}{N}\sum_{i=1}^N A_i.
\end{align*}

\noindent
The error in $\hat{\mathcal{A}}$ is
\begin{align*}
	\sigma_{\hat{\mathcal{A}}}^2 = \mathrm{E}\big[(A-\hat{\mathcal{A}})^2\big] = \frac{\sigma_A^2}{N}.
\end{align*}

\noindent
Let $B$ be another random variable for which the expected value $\mathrm{E}[B]=\mathcal{B}$ is known. By adding and subtracting, we can construct a control variate estimator for $A$:
\begin{align*}
	\hat{\mathcal{A}}_\text{cv} = \frac{1}{N}\sum_{i=1}^N\left[A_i - \alpha B_i\right] + \alpha \mathcal{B},
\end{align*}

\noindent
for any constant $\alpha$. Like $\hat{\mathcal{A}}$, this is an unbiased estimator of $\mathrm{E}[A]$. However, the error in $\hat{\mathcal{A}}_\text{cv}$ is given by
\begin{align*}
	\sigma_{\hat{\mathcal{A}}_\text{cv}}^2
						&= \frac{1}{N}\left(\sigma_A^2 + \alpha^2\sigma_B^2 - 2\alpha\,\text{cov}(A,B)\right).
\end{align*}

\noindent
Therefore, the error can be reduced if $A$ and $B$ are correlated. Differentiating, we see that the optimal value of $\alpha$ is given by
\begin{align}
	\alpha^* = \frac{\text{cov}(A,B)}{\sigma_B^2}. \label{eq:optim_alpha}
\end{align}

\noindent
For the Fermi-Dirac model considered above, $\alpha^*$ is very close to unity and we simply set $\alpha=1$ at all times. In general, the value of $\alpha^*$ could be estimated at runtime. This is useful if we add more information about the unknown variable and extend the method to a linear combination of multiple control variates (see section \ref{sec:higher}). Furthermore, the method can still be useful when a control variate is not exactly known but can be estimated more efficiently than $A$.

\begin{figure*}
	\normalsize
	\subfloat{
        \includegraphics{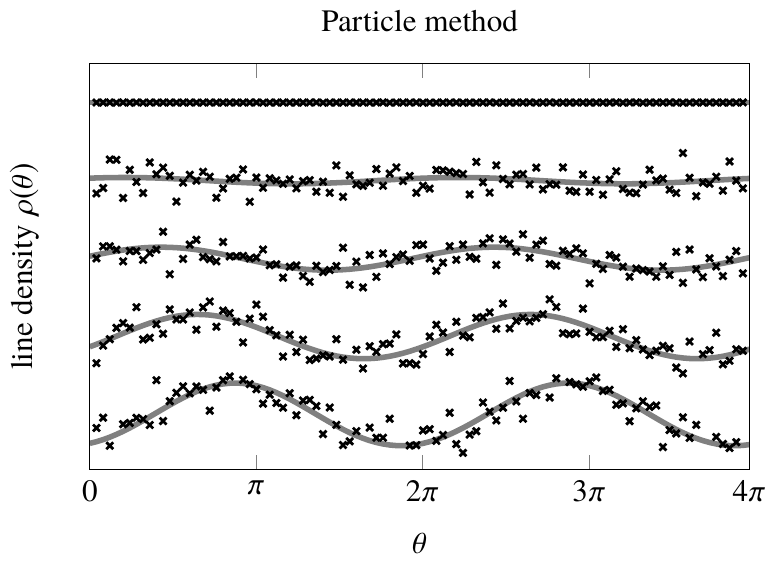}
	}\hspace{.3em}
	\subfloat{
        \includegraphics{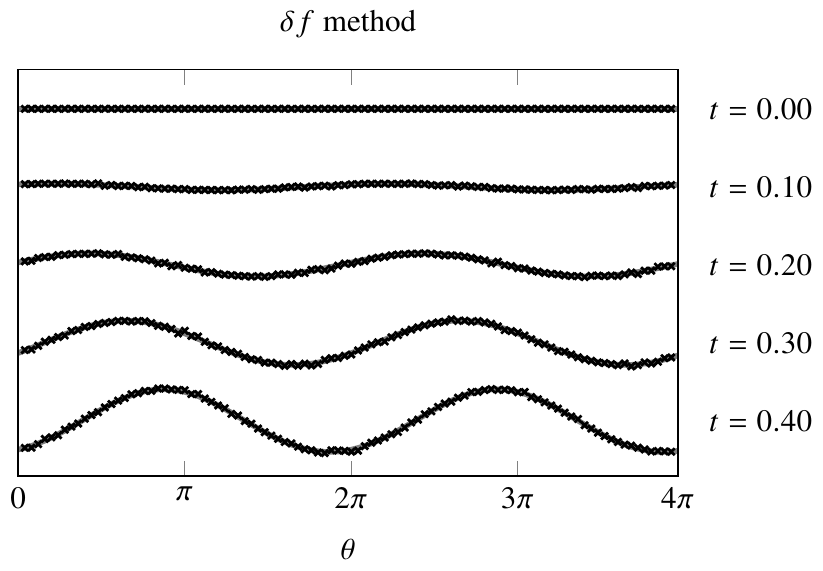}
	}
	\caption{Density profiles for the 1-dimensional elliptical sine wave test problem. We used $n=100$ position bins to create the empirical density profiles. On the left, an ordinary $N$-body simulation with $N=10^6$ particles was used. On the right, the $N$-body simulation was extended with a $\delta f$ step.}
	\label{fig:sinedens}
\end{figure*}

\subsection{Optimality}

Let us consider how the $\delta f$ method compares to other methods. To allow for the broadest possible comparison, we will write down an arbitrary hybrid method that involves some background model, $f_0(x,p,t)$, such as a fluid description or linear response, and a discrete sampling of the distribution with arbitrary particle weights, $w_i(t)$:
\begin{align}
	f_\text{hyb}&(x,p,t) = \alpha(t) f_0(x,p,t) \nonumber\\
	 &+ \sum_i w_i(t)\delta^{(3)}(x-x_i)\delta^{(3)}(p-p_i), \label{eq:hybrid_class}
\end{align}

\noindent
where $\alpha(t)$ is a weight function for the background. This parametrization captures virtually all existing methods. The ordinary $N$-body particle method corresponds to $(\alpha,w_i) = (0,1)$ at all times. Pure grid-based methods have $(\alpha,w_i)=(1,0)$. Existing hybrid methods switch over from a grid method to a particle method after some time $t_s$, which corresponds to $(\alpha,w_i) = (1-q,q)$ with $q(t) = I[t\geq t_s]$ a step function. For simplicity, we consider only the case where all particles are switched on at the same time, but the argument extends readily to the more practical case where only some particles are switched on. Given a choice of weight function, $\alpha(t)$, for the background, what choice of particle weights is optimal?

Let $f(x,p,t)$ be the nonlinear distribution and $g(x,p,t)$ the sampling distribution of the markers. In the continuous limit, the expected error in the number density is given by
\begin{align*}
	\langle\epsilon\rangle &= \int\mathrm{d}^3p\;f_\text{hyb}(x,p,t) - \int\mathrm{d}^3p\;f(x,p,t)\\
	&=\int\mathrm{d}^3p\Big(w(x,p,t)g(x,p,t) + \alpha(t)f_0(x,p,t) - f(x,p,t)\Big).
\end{align*}

\noindent
Meanwhile, the shot noise term in the power spectrum grows as the square of the particle weights, so we want to minimize
\begin{align*}
	\tfrac{1}{2}\big\langle w^2\big\rangle = \int\mathrm{d}^3p\;\tfrac{1}{2}w(x,p,t)^2g(x,p,t),
\end{align*}

\noindent
subject to the constraint $\langle\epsilon\rangle\leq\eta$ for some maximum error $\eta$. Assume that the bound is saturated. First, let us look for solutions that extremize the integral constraint. We find the unique solution
\begin{align}
	w = \frac{\delta f}{g} \;\;\;\;\;\text{with}\;\;\;\;\delta f = f - \alpha f_0. \label{eq:pure_df}
\end{align}

\noindent
This is the $\delta f$ method introduced above, with optimal $\alpha$ given by \eqref{eq:optim_alpha}. Any further solution should extremize the Lagrangian,
\begin{align*}
	\mathcal{L} = \; & \tfrac{1}{2} w(x,p,t)^2g(x,p,t)\\
	& + \lambda\Big(w(x,p,t)g(x,p,t) + \alpha(t)f_0(x,p,t) - f(x,p,t) \Big).
\end{align*}

\noindent
Writing down the Euler-Lagrange equations
\begin{align*}
	&\left[w+\lambda\right]g\nabla_p w + \tfrac{1}{2} w^2\nabla_p g + \lambda w\nabla_pg =\lambda\nabla_pf -\alpha\lambda\nabla_pf_0,
\end{align*}

\noindent
one finds a family of quadratic solutions
\begin{align*}
	w &= -\lambda \pm \sqrt{\lambda^2 + 2\lambda\frac{\delta f}{g}} \;\;\;\;\;\text{with}\;\;\;\;\delta f = f - \alpha f_0.
\end{align*}

\noindent
The case $\lambda=0$ corresponds to the trivial solution $w=0$. For $\lambda\neq0$, we obtain the minima
\begin{align}
	 w &= \frac{\delta f}{g} - \frac{1}{2\lambda}\left(\frac{\delta f}{g}\right)^2 + \mathcal{O}\left(\frac{1}{\lambda^{2/3}}\frac{\delta f}{g}\right)^3. \label{eq:deltaf_methods}
\end{align}

\noindent
These solutions correspond to small perturbations around the $\delta f$ method that trade some accuracy for a possible reduction in shot noise. However, since the leading correction is $\propto(\delta f)^2$, this is only possible if the background model is skewed with respect to the nonlinear solution. Typically, the skewness and the additional reduction in shot noise is negligible. In fact, since the next-to-leading correction is positive, shot noise increases if the skewness is small.

We have shown that within the broad class of hybrid methods described by equation \eqref{eq:hybrid_class}, $\delta f$-type methods of the form \eqref{eq:deltaf_methods} minimize the amount of shot noise, subject to the constraint that the error in the number density remains below a certain bound. The $\delta f$ method given by \eqref{eq:pure_df}, recovered from \eqref{eq:deltaf_methods} in the limit $\lambda\to\infty$, is the unique solution for which the expected error $\langle\epsilon\rangle=0$. The optimal value of $\alpha$ is given by \eqref{eq:optim_alpha}, but will be close to 1 if $f_0\approx f$. This is the method we will use exclusively, with the choice $\alpha=1$.

\section{One-dimensional example}\label{sec:verification}

We now illustrate the method by applying it to a one-dimensional test problem with a known solution. Readers that are satisfied with the mathematical derivation may skip ahead to section \ref{sec:relativity}.

\subsection{The elliptical sine wave}\label{sec:sinewave}

Consider the 1-dimensional collisionless Boltzmann equation
\begin{align*}
	\frac{\partial f}{\partial t} + p\frac{\partial f}{\partial x} - \frac{\partial\Phi}{\partial x}\frac{\partial f}{\partial p} = 0,
\end{align*}

\noindent
where the particles move under a conservative force $F(x)=-\Phi'(x)$. Let us assume a periodic potential given by
\begin{align*}
	\Phi(x) = \sin^2(x/2).
\end{align*}

\noindent
The steady-state solution can be found to be:
\begin{align*}
	f(x,p) = \frac{\rho_0}{\sqrt{2\pi\sigma^2}}\exp\left(-\frac{p^2}{2\sigma^2}+\frac{\cos(x)}{2\sigma^2}\right),
\end{align*}

\noindent
in terms of the background density $\rho_0$ and velocity dispersion $\sigma$. The corresponding density profile $\rho(x)$ is given by
\begin{align*}
	\rho(x) &= \int_{-\infty}^\infty f(x,p)\mathrm{d}p= \rho_0 \exp\left(\frac{\cos(x)}{2\sigma^2}\right).
\end{align*}

\noindent
To find the general time-dependent solution, we use the method of characteristics. The characteristic equations are
\begin{align*}
	\frac{\mathrm{d}x}{\mathrm{d}t} = p,\;\,\;\,\;\,\;\,\;\,\;\,\;\,\;\,\;\,\;\,\frac{\mathrm{d}p}{\mathrm{d}t} = -\frac{1}{2}\sin(x).
\end{align*}

\noindent
These equations of motion can be solved in terms of the energy $E=\tfrac{1}{2}p^2+\sin^2(x/2)$, which gives
\begin{align*}
	\sin (x/2) = \mathrm{sn}\left(\pm\sqrt{E/2}(t - t_0)\right),
\end{align*}

\noindent
where $t_0$ is an integration constant and $\mathrm{sn}(x)$ is the Jacobi elliptic sine function with elliptic modulus $k=1/\sqrt{E}$ \citep{weisstein02}\footnote{For $E\to\infty$, we have $\mathrm{sn}\,x\to\sin x$, meaning that $x\propto t$. The particle `ignores' the potential. For $E=k=1$, $\mathrm{sn}\,x=\tanh x$, meaning the particle asymptotically approaches a potential peak. For $E<1$, the particle is bounded and oscillates between peaks.}. Assuming a homogeneous Gaussian distribution with mean $\bar{p}$ for the initial momenta $p_0$,
\begin{align}
	f(x_0,p_0,t_0) = \frac{\rho_0}{\sqrt{2\pi\sigma^2}}\exp\left(-\frac{(p_0-\bar{p})^2}{2\sigma^2}\right), \label{eq:sine_ini}
\end{align}

\noindent
the general solution, $f(x,p,t)$, at later times is a complicated expression involving elliptic sines and arcsines. The details are given in Appendix~\ref{sec:sinedetails}. We replicate the problem using $N$-body methods. A large number of particles are initialized on the interval $x\in[0,4\pi]$ with momenta drawn from the initial distribution \eqref{eq:sine_ini}. The particles are then integrated using
\begin{align*}
	\Delta x = p\Delta t, \;\,\;\,\;\,\;\,\;\,\;\,\;\,\;\,\;\,\;\,       \Delta p = -\frac{1}{2}\sin(x)\Delta t.
\end{align*}

\noindent
In addition to the ordinary $N$-body method, we use a $\delta f$ method, where the background model is given by
\begin{align*}
	f_0(x,p,t) = \frac{\rho_0}{\sqrt{2\pi\sigma^2}}\exp\left(-\frac{(p-\bar{p})^2}{2\sigma^2}\right),
\end{align*}

\noindent
and the weights are updated during each step via $w=\delta f/f$. The corresponding density profiles are shown in Fig.~\ref{fig:sinedens}. The plots were created using $N=10^6$ particles and the model parameters are $\rho_0=\sigma=1$ and $\bar{p}=10$. The results show that both the ordinary $N$-body simulation and the simulation with a $\delta f$ step can reproduce the exact solution. However, the ordinary method is very noisy, whereas the $\delta f$ method reproduces the expected profiles with remarkable accuracy. The reason for this is that while the distribution itself has a large dispersion, resulting in noisy results for the ordinary method, the perturbations from the steady solution are small, which allows the $\delta f$ method to work. This is exactly analogous to the cosmic neutrino background.

\section{Relativistic effects}\label{sec:relativity}

Neutrinos constitute a relativistic fluid at early times, which introduces some subtleties when evolving such a fluid with a Newtonian code. Including relativistic effects is not necessary for the $\delta f$ method, but we include them in our simulations to allow for a consistent comparison with recent works \citep{adamek17,tram19,partmann20}. Furthermore, the higher order $\delta f$ methods discussed in section \ref{sec:higher} provide a natural setting for including these effects without neglecting the nonlinear evolution of the neutrinos. We will work in the Newtonian motion framework of \citet{fidler17} and make modifications to the initial conditions, long-range force calculation, and particle equations of motion as outlined below.

\begin{figure*}
	\normalsize
	\centering
	\subfloat{
	\includegraphics{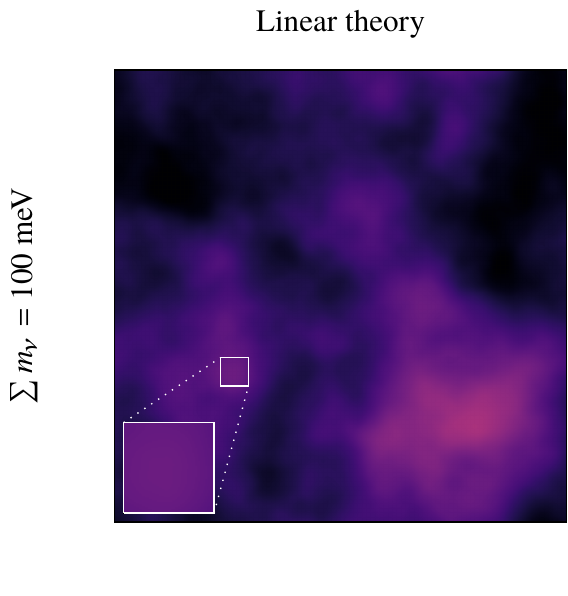}
	}\hspace{-3em}
	\subfloat{
        \includegraphics{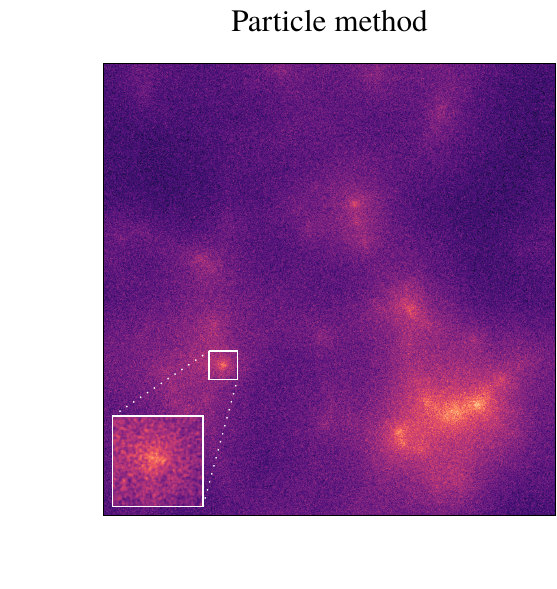}
	}\hspace{-3em}
	\subfloat{
        \includegraphics{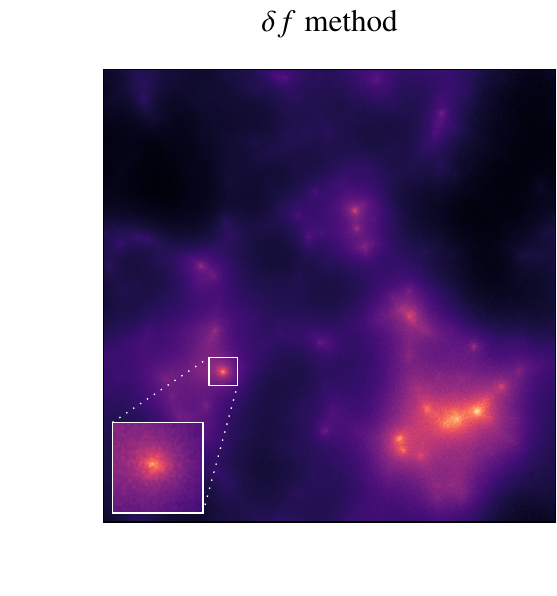}
	}\\\vspace{-5.5ex}
	\subfloat{
        \includegraphics{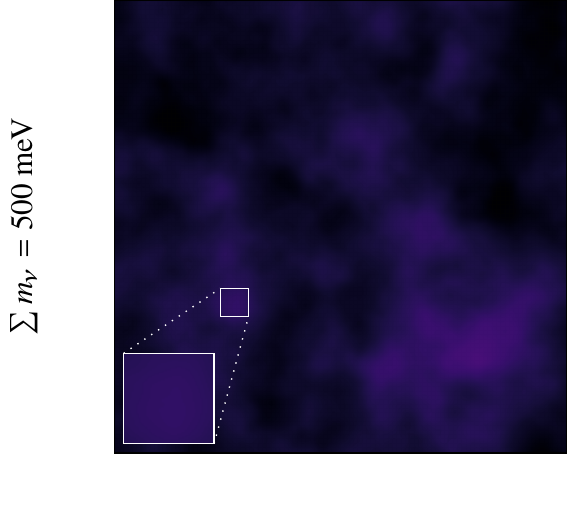}
	}\hspace{-3em}
	\subfloat{
        \includegraphics{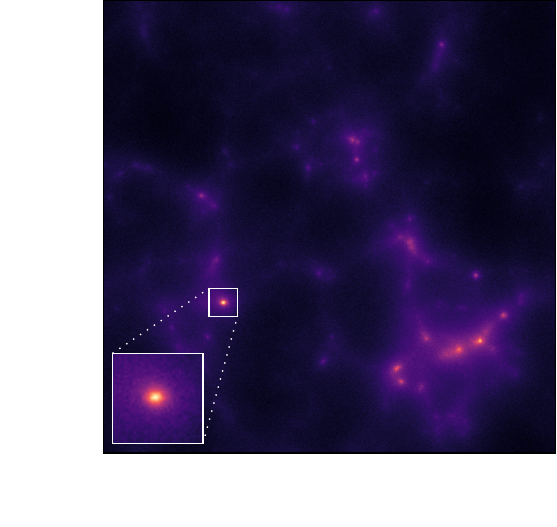}
	}\hspace{-3em}
	\subfloat{
        \includegraphics{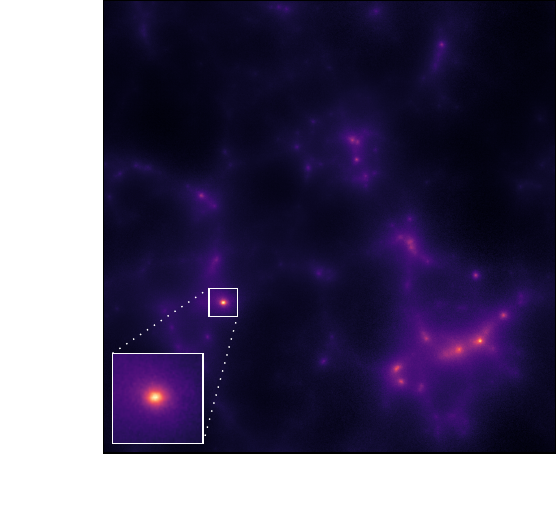}
	}
	\caption{Neutrino density plots of $(256\text{ Mpc})^3$ cubes at $z=0$, simulated with two commonly-used methods and with the $\delta f$ method. The particle and $\delta f$ simulations used $N_\nu=1024^3$ particles. Shot noise is clearly visible for the particle method, although noticably less so for $\sum m_\nu=500\text{ meV}$. The linear theory model fails to reproduce the small-scale behaviour. The $\delta f$ method solves both problems. The inset zooms in on a neutrino halo, which should be compared with the linear response prediction in Fig.~\ref{fig:density_plots2}.}
	\label{fig:density_plots}
\end{figure*}

\subsection{Initial Conditions}

To generate initial conditions for massive neutrinos and to set up the higher order background models (section \ref{sec:higher}), accurate calculation of the linear theory neutrino distribution function is indispensable. This can be done with the Boltzmann codes \textsc{camb} \citep{lewis11} and \textsc{class} \citep{lesgourgues11}. At their default settings, these codes produce accurate total matter and radiation power spectra (their intended purpose), but the neutrino related transfer functions (e.g. density and velocity) are  not converged and can be very inaccurate \citep{dakin19}. To obtain converged results, we post-process perturbation vectors from \textsc{class} by integrating source functions up to multipole $\ell_\text{max}=2000$. This prevents the artificial reflection that can happen for low $\ell_\text{max}$. See appenfix \ref{sec:firebolt} for more details.

Initial conditions are then created using the post-processed transfer functions from \textsc{class} in $N$-body gauge at $z=100$. We do not follow the usual approach of back-scaling the present-day power spectrum, but use the so-called forward Newtonian motion approach \citep{fidler15,fidler17}. To our knowledge, forward Newtonian motion initial conditions have always been set up with the Zel'dovich approximation. However, this approximation is known to be inadequate for precision simulations \citep{crocce06}. To go beyond Zel'dovich initial conditions, we determine the Lagrangian displacement vectors $\psi=x-q$ by solving the Monge-Amp\`ere equation
\begin{align*}
	\rho(x) = \bar{\rho}(1+\delta(x)) = \left\rvert 1 + \frac{\partial\psi(x)}{\partial q}\right\rvert.
\end{align*}

\noindent
This equation is solved numerically with a fixed-point iterative algorithm that exploits the fact that the density perturbation $\delta$ is small. We note that this approach is not equivalent to Lagrangian perturbation theory, but merely provides a more accurate map from the Eulerian initial density field to a Lagrangian displacement field compared to the Zel'dovich approximation. A detailed analysis of this method will be presented elsewhere. Velocities were determined independently using the transfer functions for the velocity dispersion $\theta=ik\cdot v$. Neutrino particles were displaced randomly in phase space according to the perturbed phase-space density function, $f_\text{PT}(x,p,t)$, including terms up to $\ell=5$.

\subsection{Long-range forces}

In the weak-field limit, working in $N$-body gauge, the continuity and Euler equations for a collisionless fluid can be written as \citep{fidler15,fidler17b,brandbyge17}:
\begin{align*}
	\dot{\delta} + \nabla\cdot\mathbf{v} &= 0,\\
	\dot{\mathbf{v}} + aH\mathbf{v} &= -\nabla\phi + \nabla\gamma^\text{Nb},
\end{align*}

\noindent
where overdots denote conformal time derivatives, $\delta$ is the density contrast, $\mathbf{v}$ the peculiar velocity, and $H=\dot{a}/a^2$. The scalar potential $\phi$ receives contributions from all fluid components:
\begin{align*}
	\nabla^2\phi = 4\pi Ga^2\sum_{i}\delta\rho_i,
\end{align*}

\noindent
where the sum runs over cold dark matter, baryons, neutrinos, and photons. The $N$-body gauge term, $\nabla\gamma^\text{Nb}$, arises from the anisotropic stress of relativistic species. In the absence of such species, the continuity and Euler equations agree with the Newtonian equations solved in conventional $N$-body codes. This is what makes $N$-body gauge useful as it allows one to set up initial conditions in $N$-body gauge, evolve them in a Newtonian simulation, and give the results a relativistic interpretation. The relativistic corrections become relevant at the 0.5\% level on the largest scales in our Gpc simulations. We include the contribution from photons and massless neutrinos (and for some runs, the massive neutrinos\footnote{Specifically, the linear theory runs and the runs with higher-order $\delta f$ methods, as discussed in sections \ref{sec:results} and \ref{sec:higher}, respectively.}) by realizing the corresponding transfer functions from \textsc{class} on a grid as part of the long-range force calculation in our $N$-body code \textsc{swift}.

\subsection{Particle content}

When simulating light neutrinos from high redshifts, we are evolving relativistic particles in a Newtonian simulation. Such particles can reach superluminal speeds when evolved using the ordinary equations of motion. Following \citet{adamek16}, we address this issue by replacing the equations of motion with semi-relativistic equations that are valid to all orders in $u$:
\begin{align*}
	\dot{\mathbf{u}} &= - \frac{2u^2 + a^2}{\sqrt{u^2 + a^2}}\nabla\left(\phi-\gamma^\text{Nb}\right),\\
	\dot{\mathbf{x}} &= \frac{\mathbf{u}}{\sqrt{u^2 + a^2}}.
\end{align*}

\noindent
Here, $a^{-1}\mathbf{u}$ is the spatial part of the 4-velocity. Additionally, when computing the energy density, we replace the weighted mass of the particles with a weighted energy $\epsilon_i=\sqrt{m^2+p_i^2}$. We have verified that both changes result in virtually identical power spectra at $z=0$, but with the added benefit of slightly reducing the number of timesteps needed to integrate the particles.

\begin{table}
	\centering
	\caption{Description of the simulations. The listed particle mass, $m_p$, refers to the combined cold dark matter and baryon particles.}
	\begin{tabular}{  l  l  l  l  l }
		\hline
		\hline
		Side Length & $N_c$ & $m_p\;\big[\,M_\odot\,\big]$ & $N_\nu$ & $\sum m_\nu$ \\
		\hline
		1024 Mpc & $1024^3$ & $3.96\times10^{10}$ & 0 & 0 meV \\
		1024 Mpc & $1024^3$ & $3.93\times10^{10}$ & $1024^3$ & 100 meV \\
		1024 Mpc & $1024^3$ & $3.81\times10^{10}$ & $1024^3$ & 500 meV \\
		256 Mpc & $512^3$ & $4.95\times10^{9}$ & 0 & 0 meV \\
		256 Mpc & $512^3$ & $4.92\times10^{9}$ & $1024^3$ & 100 meV\\
		256 Mpc & $512^3$ & $4.77\times10^{9}$ & $1024^3$ & 500 meV \\
		\hline
		\hline
	\end{tabular}
	\label{tab:sims}
\end{table}

\begin{figure*}
	\normalsize
	\centering
	\subfloat{
        \includegraphics{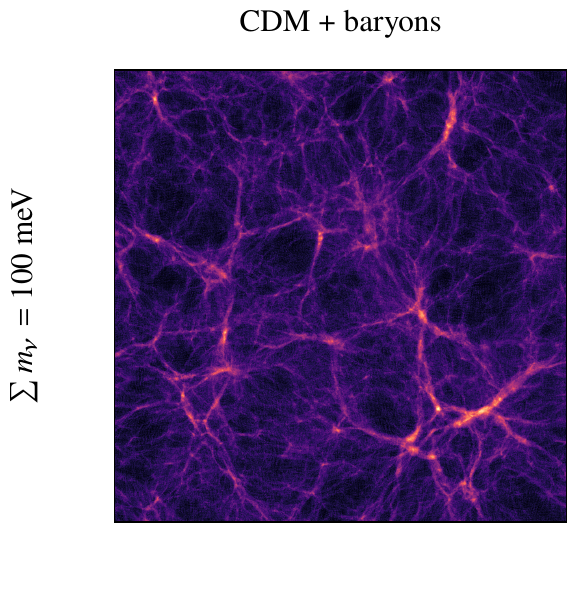}
	}\hspace{-3em}
	\subfloat{
        \includegraphics{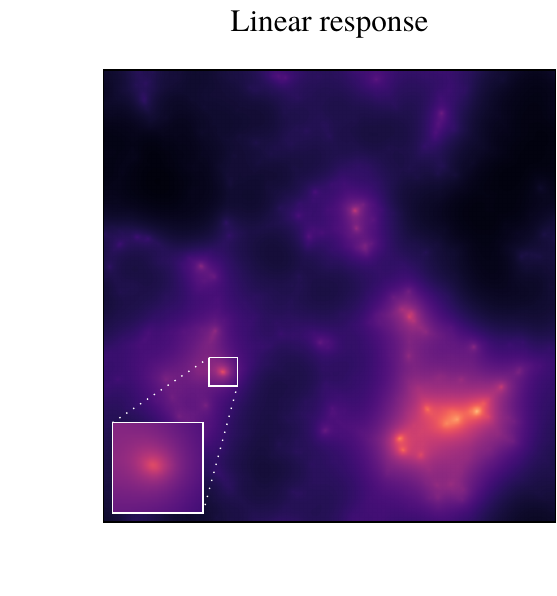}
	}\hspace{-3em}
	\subfloat{
        \includegraphics{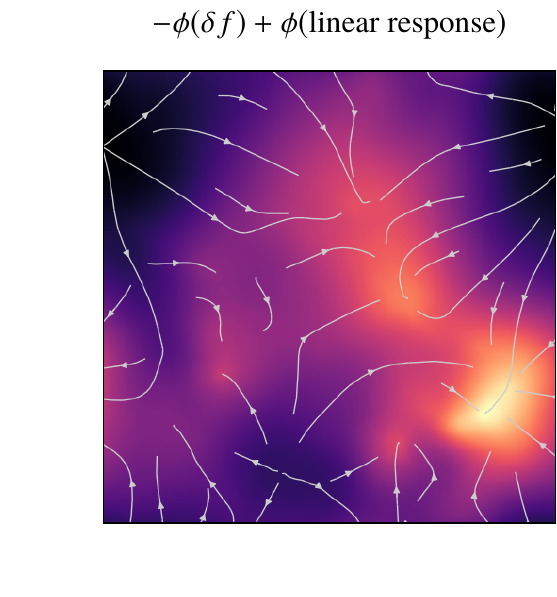}
	}\\\vspace{-5.5ex}
	\subfloat{
        \includegraphics{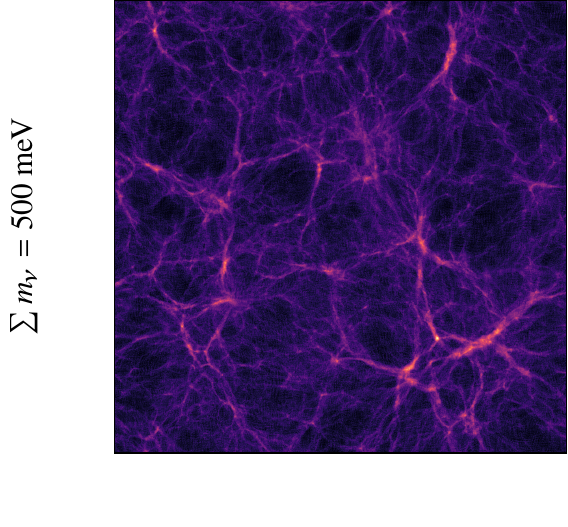}
	}\hspace{-3em}
	\subfloat{
        \includegraphics{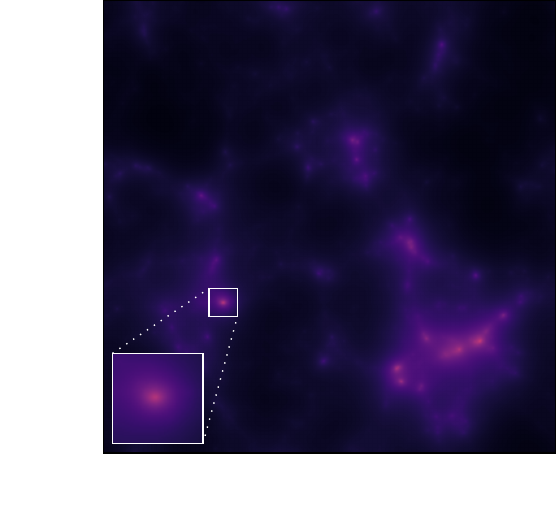}
	}\hspace{-3em}
	\subfloat{
        \includegraphics{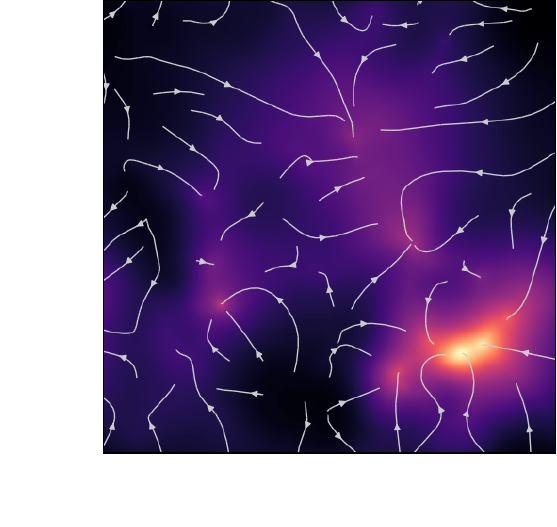}
	}
	\caption{Density plots of $(256\text{ Mpc})^3$ cubes at $z=0$. The linear response method applies the linear theory ratio $\delta^\text{lin}_\nu(k)/\delta^\text{lin}_\text{cb}(k)$ to the simulated cdm + baryon phases \citep{ali12}. Compared to the linear theory prediction, it performs remarkably well, but the neutrino halos around clusters are significantly more diffuse compared to the particle and $\delta f$ simulations (compare the zoomed in halo with the $\delta f$ prediction in Fig.~\ref{fig:density_plots}). The resulting potential difference is shown in the last column, with flowlines indicating the forces that are not present in the linear response model.}
	\label{fig:density_plots2}
\end{figure*}

\section{Simulations}\label{sec:sims}

We now describe our neutrino simulations, which were run on the \textsc{cosma6} computing facility in Durham. We have implemented the $\delta f$ method in the cosmological hydrodynamics code \textsc{swift} \citep{schaller16,schaller18}. \textsc{swift} uses a task-based parallelization paradigm to achieve strong scaling on large clusters and obtain significant speed-ups over competing $N$-body codes. The main simulations presented in this paper use the basic version of the $\delta f$ method with a homogeneous Fermi-Dirac distribution as background model. Our choice of cosmological parameters, based on Planck 2018 \citep{planck18}, is $(h, \Omega_\text{c}+\Omega_{\nu}, \Omega_\text{b}, A_s, n_s)=(0.6737, 0.265, 0.0492, 2.097\times10^{-9}, 0.9652)$. We run two sets of simulations at different resolution to test the large-scale and small-scale behaviour of various methods. The cube sizes and particle numbers are listed in Table~\ref{tab:sims}.

\subsection{Choice of neutrino masses}\label{sec:numass}

Neutrino oscillations indicate that there are three neutrino mass eigenstates with unknown masses $m_i$. The mass splittings have been measured with good precision to be
\begin{align*}
\Delta m_{21}^2&\equiv m_2^2 - m_1^2=7.42_{-0.20}^{+0.21}\times 10^{-5}\text{ eV}^2,\\
\Delta m_{3\ell}^2&\equiv m_3^2-m_\ell^2=\begin{cases}
+2.514_{-0.027}^{+0.028}\times 10^{-3}\text{ eV}^2 \text{ (NO)},\\
-2.497_{-0.028}^{+0.028}\times 10^{-3}\text{ eV}^2 \text{ (IO)}.\end{cases}
\end{align*}

\noindent
The sign of $\Delta m_{21}^2$ is known to be positive, which leaves two possible mass orderings: $m_1 < m_2 < m_3$ with $\ell=1$ (normal) or $m_3 < m_1 < m_2$ with $\ell=2$ (inverted). Current oscillation data slightly favour the normal ordering at $1.6\sigma$  \citep{esteban20}.

The best terrestrial constraint on the absolute mass scale comes from the \textsc{katrin} detector, which places a bound of $m_\beta<1.1\text{ eV}$ at the 90\% C.L. on the effective neutrino mass \citep{aker19}. Assuming a quasi-degenerate mass spectrum, this corresponds to a neutrino mass sum of $\sum m_\nu\approx 3\text{ eV}$. Recent cosmological limits are much stronger and are quoted below at the 95\% C.L. Assuming a degenerate mass spectrum, the Planck temperature, polarization, and lensing likelihoods give a constraint of $\sum m_\nu<0.24\text{ eV}$ or $\sum m_\nu<0.26\text{ eV}$, depending on the details of the high-$\ell$ polarization analysis \citep{planck18}. Adding BAO data from BOSS DR12, MGS, and 6dFGS, \citet{choudhury20} found $\sum m_\nu<0.12\text{ eV}$ (degenerate), $\sum m_\nu<0.15$ eV (normal), and $\sum m_\nu<0.17$ eV (inverted). An analysis of the shape of the BOSS DR11 redshift-space power spectrum, combined with CMB data and Type 1a supernovae leads to $\sum m_\nu<0.18\text{ eV}$ \citep{upadhye19}. Adding instead the latest SDSS DR14 BOSS and eBOSS Lyman-$\alpha$ forest data to the Planck and BAO data leads to the strongest constraint: $\sum m_\nu<0.09$ eV \citep{palanque20}.

Given these limits, we consider three values for $\sum m_\nu$, keeping the present-day value $\Omega_{\text{m},0} = \Omega_{\text{cb},0} + \Omega_{\nu,0}$ fixed. Scenario one contains three massless neutrinos, scenario two corresponds to the inverted mass ordering with $\sum m_\nu=100$ meV\footnote{Specifically, two $0.0486$ eV neutrinos and one massless neutrino.}, and scenario three to a degenerate spectrum with $\sum m_\nu=500$ meV. The first two models bracket the most interesting range of values $0<\sum m_\nu<100$ meV. The last model has surely been ruled out, but is included for several reasons. First of all, the $\delta f$ method reduces to the ordinary particle method in the large mass limit at late times. Hence, the $\sum m_\nu=500$ meV case provides a useful consistency check. Second, when simulations are used to emulate statistics for parameter extraction, we should allow for unlikely excursions in MCMC analyses without our simulation methods breaking down \citep{partmann20}. Finally, in the extended parameter space around $\Lambda$CDM, for example with a non-standard lensing amplitude, $A_L$, or curvature, or when varying the dark energy equation of state, the possibility of larger neutrino masses remains very relevant \citep{mccarthy18,valentino20,upadhye19,choudhury20}.

The two massive scenarios considered in this paper have degenerate neutrino masses ($2\times50$ meV and $3\times 167$ meV). However, the $\delta f$ method can easily be extended to account for mass splittings. In that case, particles would be labelled with a given mass state, $i$, and each state would have its own background model, $f_{0,i}$. The reduction in shot noise is largest for the smallest neutrino masses, placing different masses on a level footing. This allows for better load balancing between different neutrino masses.

\section{Results}\label{sec:results}

We compare our neutrino $\delta f$ method with three commonly used alternatives. The most common alternative is the ordinary $N$-body particle method, which is the same in every respect as our method, but with the weighting step disabled. Next, we consider a linear theory method based on \citet{tram19} that does not evolve neutrino particles but instead realizes the linear theory neutrino perturbation in $N$-body gauge on a grid. The neutrinos are then fully accounted for in the long-range forces. Finally, we consider the linear response method of \citet{ali12} in which the neutrino perturbation is calculated by applying the linear theory transfer function ratio, $\delta^\text{lin}_\nu(k)/\delta^\text{lin}_\text{cdm+b}(k)$, to the simulated cdm + baryon phases.

A visual inspection of the neutrino density plots shown in Figs.~\ref{fig:density_plots} and~\ref{fig:density_plots2} reveals the strengths and weaknesses of the four methods. Broadly, we see that the linear theory method does not suffer from shot noise, but fails to reproduce the small-scale behaviour resolved by the particle and $\delta f$ methods. At the same time, shot noise is clearly visible in the particle simulation with $\sum m_\nu=100$ meV, despite using $N_\nu=1024^3$ particles in a $256\text{ Mpc}$ cube. This is evidently cured in the $\delta f$ plot. We also see that shot noise is much less of a problem for $\sum m_\nu=500$ meV, but  the $\delta f$ plot is still less grainy than the corresponding particle plot. Finally, Fig.~\ref{fig:density_plots2} shows that the linear response method greatly improves on the pure linear theory prediction, but still produces neutrino halos that are too diffuse compared to the particle and $\delta f$ simulations.

\begin{figure*}
	\normalsize
	\subfloat{
        \includegraphics{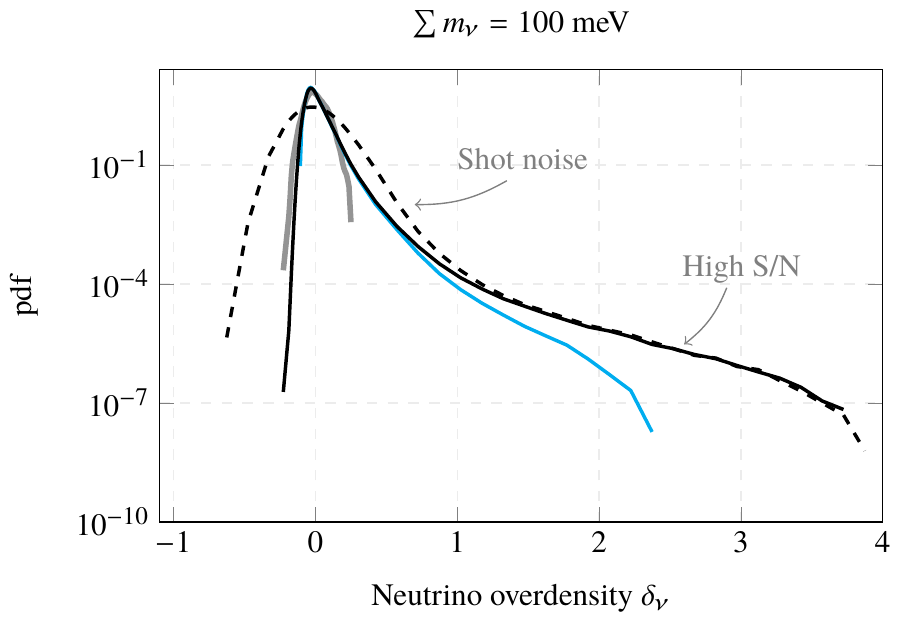}
	}
	\subfloat{
        \includegraphics{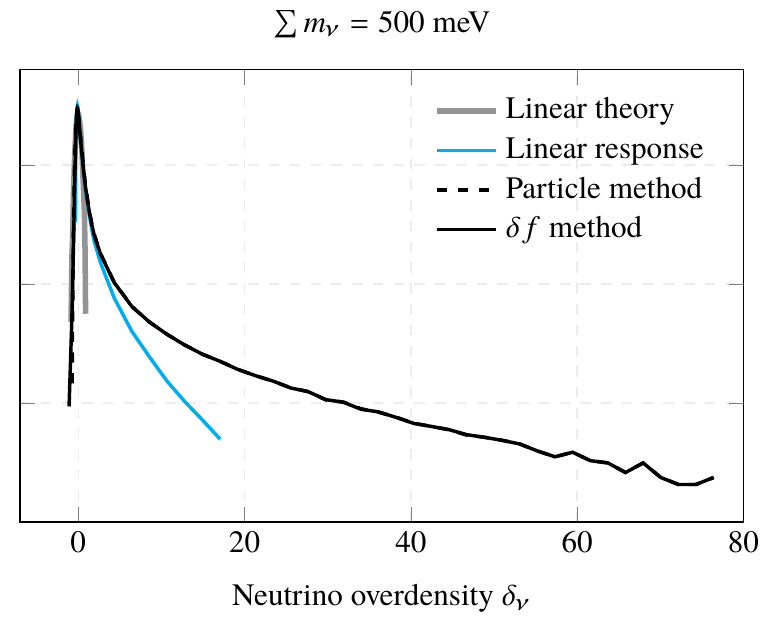}
	}

	\caption{Neutrino probability density functions (pdf) at $z=0$, computed on a $1024^3$ grid from the 256 Mpc simulations, and smoothed with a Gaussian filter with radius $R=256$ kpc. We compare the $\delta f$ method with three commonly used alternatives. The particle and $\delta f$ methods agree in the high density tail, because the largest overdensities have enough particles to achieve a high signal-to-noise ratio. Shot noise plagues the particle method, particularly in underdense regions. The linear methods fail in the high density tail.}
	\label{fig:nupdf}
\end{figure*}

\subsection{Neutrino component}

We start with an analysis of the probability density function of the neutrino density field, computed on a $1024^3$ grid from the 256 Mpc simulations. Refer to the plots in Fig.~\ref{fig:nupdf}, which bear out the basic picture sketched above. For the $\sum m_\nu=100$ meV neutrinos, the particle method is plagued by shot noise, but agrees with the $\delta f$ method in the high density tail where the particle number is sufficient to obtain a good signal-to-noise ratio. The linear prediction fails in the high and low density tails. Finally, the linear response method, which applies the linear theory ratio $\delta_\nu(k)/\delta_\text{cdm+b}(k)$ to the cdm+baryon phases, is an intermediate case between the linear theory and $\delta f$ methods. For the more massive scenario, the situation is much the same, except that shot noise is much less of a problem for the particle method on these scales.

\begin{figure*}
 	\normalsize
 	\subfloat{
        \includegraphics{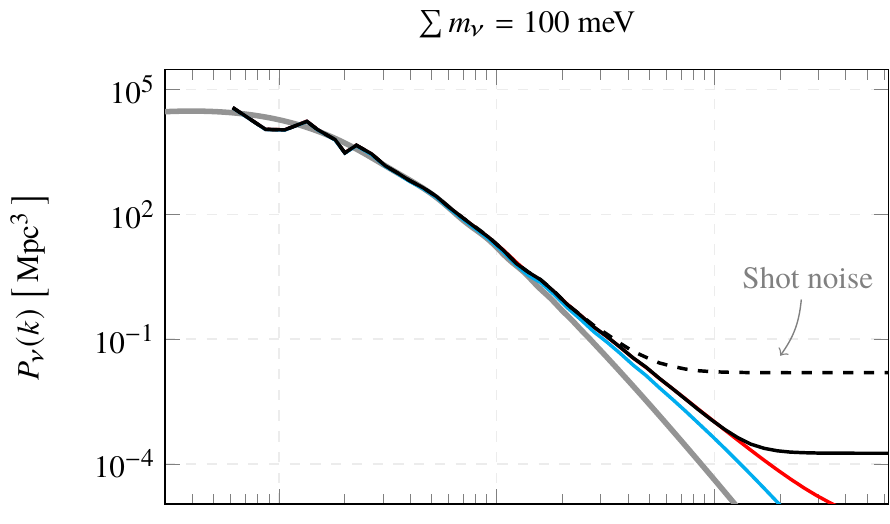}
 	}
 	\subfloat{
        \includegraphics{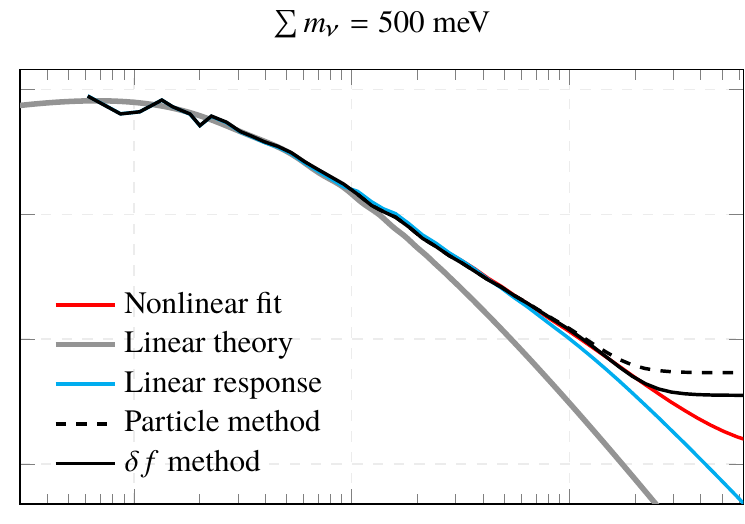}
 	}\\\vspace{-1em}
 	\subfloat{
        \includegraphics{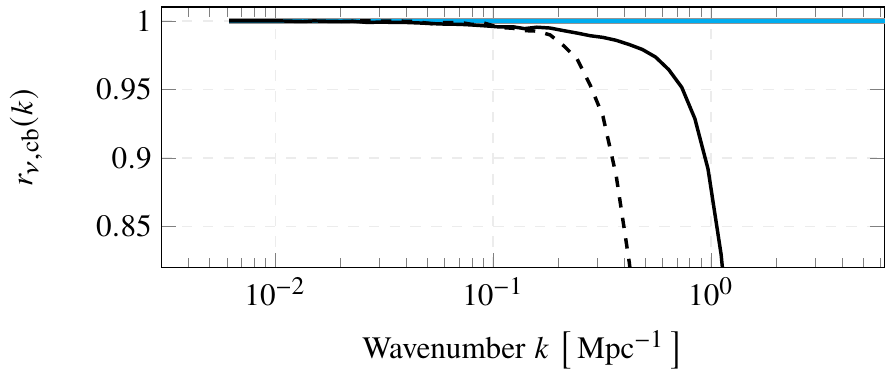}
 	}
    \subfloat{
    	\includegraphics{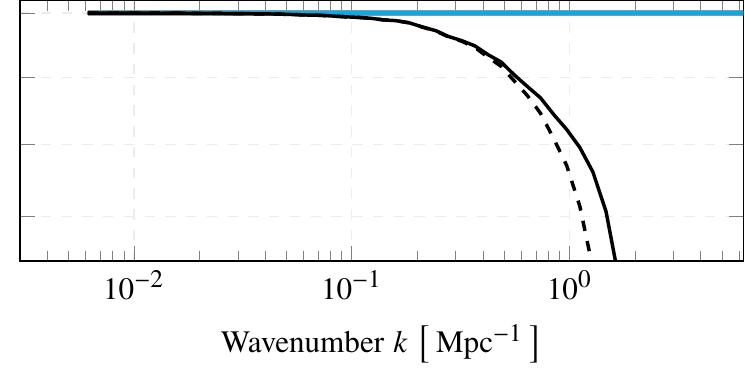}
 	}

 	\caption{Neutrino power spectra at $z=0$. We compare the $\delta f$ method with three commonly used alternatives. Shot noise enters the power spectrum at the constant level $V/N=1/64$ for the particle method. We also show a fit to the $\delta f$ power spectrum (red curves), given by eq. \eqref{eq:nufit}. The bottom panels show the cross-spectral coefficient $r_{\nu,\text{cb}}=P_{\nu,\text{cb}}/\sqrt{P_\nu P_\text{cb}}$.}
 	\label{fig:nupower}
 \end{figure*}

Next, we consider two-point statistics and show the neutrino power spectrum at $z=0$ in Fig.~\ref{fig:nupower}, combining the large and small simulations to show a wide range of scales. We use the Gpc simulations for $k<0.1\text{ Mpc}^{-1}$ and the 256 Mpc simulations for $k\geq 0.1\text{ Mpc}^{-1}$. As expected, all methods agree on scales greater than $k=0.1\;\text{Mpc}^{-1}$ for both neutrino masses. On smaller scales, linear theory significantly underpredicts the amount of neutrino clustering. The linear response method also underpredicts the neutrino power spectrum, but not by as much. The relative difference between the nonlinear power spectrum and linear power spectrum is greater for neutrinos than for cdm and baryons. To account for this effect, we fit a nonlinear correction to the linear response power spectrum using the measured $\delta f$ power spectrum up to $k=1\;\text{Mpc}^{-1}$:
\begin{align}
	P^\text{fit}_\nu(k) = P_\text{cdm+b}(k)\left[\frac{\delta^\text{lin}_\nu(k)}{\delta^\text{lin}_\text{cdm+b}(k)}\right]^2e^{\alpha + \beta k},\label{eq:nufit}
\end{align}

\noindent
and find $\alpha=0.006\pm0.004$ and $\beta=0.90\pm0.01$ ($\sum m_\nu = 100$ meV) and $\alpha=-0.06\pm0.03$ and $\beta=0.34\pm0.09$ ($\sum m_\nu = 500$ meV). These are shown as the red curves in Fig.~\ref{fig:nupower}.

The particle simulations are clearly affected by shot noise, at the level of $V/N=1/64$, obscuring the neutrino signal on scales smaller than $k=0.2\;\text{Mpc}^{-1}$ for the lightest scenario and on scales smaller than $k=1\;\text{Mpc}^{-1}$ for the more massive scenario. Using the $\delta f$ method, shot noise is significantly reduced in the former case (factor of 87) and slightly reduced in the latter case (factor of 3.5), revealing a signal down to $k=1-2\;\text{Mpc}^{-1}$. Hence, $\delta f$ simulations can achieve a similar resolution independently of mass without adjusting the particle number.

We also show the cross-spectral coefficient
\begin{align*}
	r_{\nu, \text{cb}}(k) = \frac{P_{\nu,\text{cb}}(k)}{\sqrt{P_\nu(k)P_\text{cb}(k)}},
\end{align*}

\noindent
which captures phase differences between the dark matter and neutrinos. By definition, $r_{\nu,\text{cb}}=1$ according to the linear response method. However, this does not hold on small scales as can be seen in the bottom panels. Up to the point where shot noise becomes a problem, the particle and $\delta f$ methods agree, demonstrating that $r_{\nu,\text{cb}}<1$. This is particularly clear for $\sum m_\nu=500$ meV.

\begin{figure*}
	\normalsize
	\subfloat{
        \includegraphics{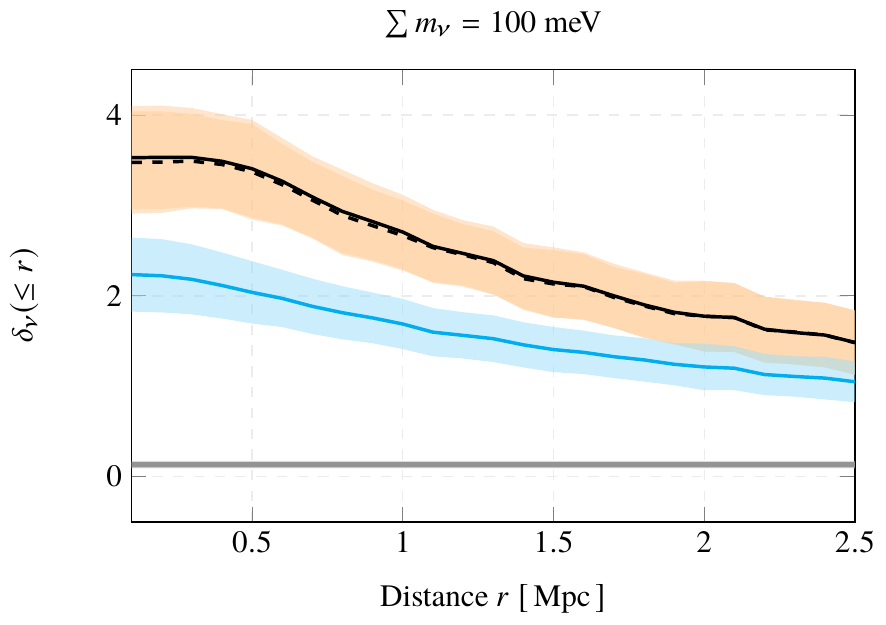}
	}
	\subfloat{
        \includegraphics{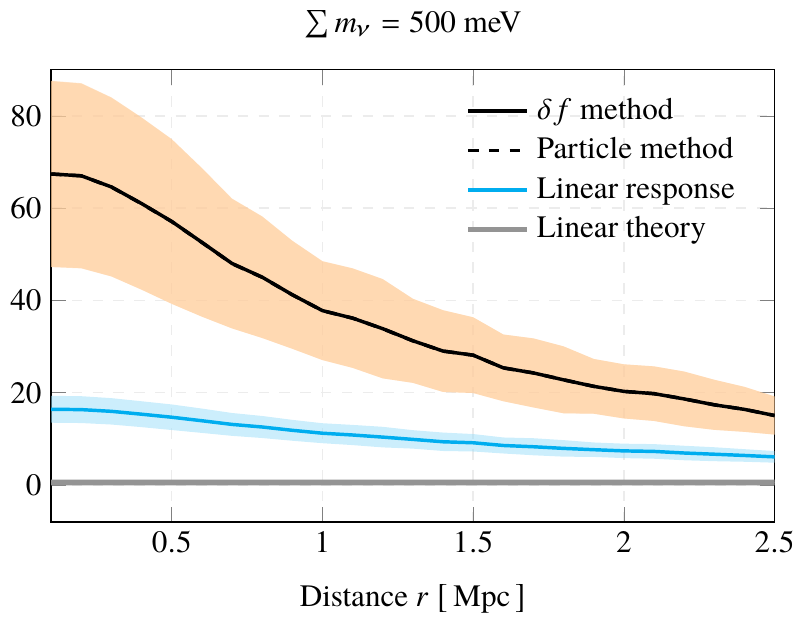}
	}

	\caption{Stacked neutrino density profiles at $z=0$ for halos with virial mass $M_\text{cdm+b}$ in the range $(5,14)\times 10^{14}\,M_\odot$, computed with four different methods from the 256 Mpc simulations. The particle and $\delta f$ curves overlap almost perfectly. The shaded area indicates the $1\sigma$ dispersion around the mean profile.}
	\label{fig:halo_profiles}
\end{figure*}

Next, we consider how well the simulations can resolve the extended neutrino halos surrounding galaxies and clusters \citep{brandbyge10b,villaescusa11}. In Fig.~\ref{fig:halo_profiles}, we show stacked neutrino profiles for halos with virial mass $M_\text{cdm+b}$ in the range $(5,12)\times10^{14}M_\odot$. The particle and $\delta f$ methods agree almost perfectly, once again because of the high signal-to-noise ratio in the largest overdensities. In linear theory, the neutrino halos are completely absent as is evident also from the cross-sections in Fig.~\ref{fig:density_plots}. Finally, the linear response method predicts neutrino halos that are too diffuse compared to the nonlinear simulations, and with too little dispersion from the mean profile. The larger dispersion found in the nonlinear simulations is not due to errors in individual profiles, but due to a stronger correlation between $M_\text{cdm+b}$ and the local neutrino density.

\begin{figure*}
	\normalsize
	\subfloat{
		\includegraphics{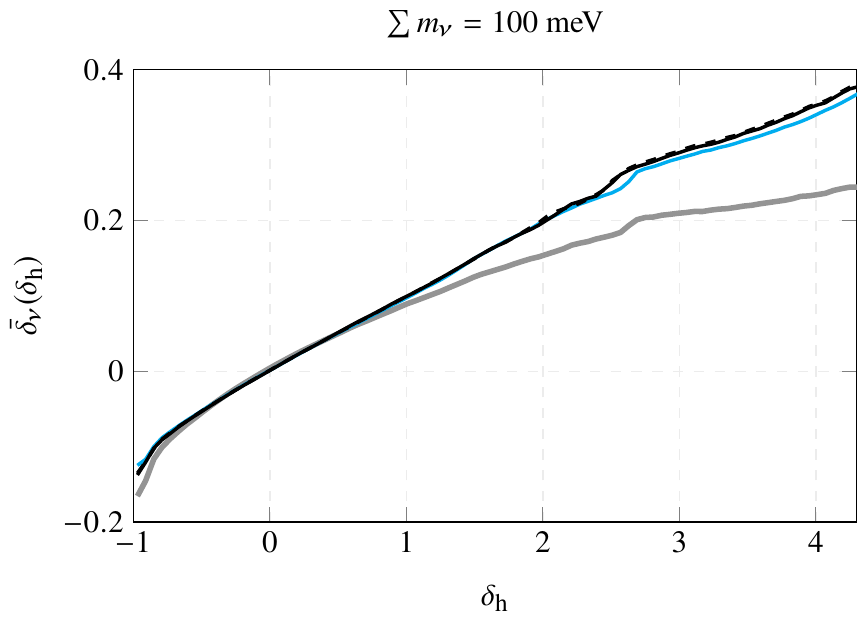}
	}
	\subfloat{
		\includegraphics{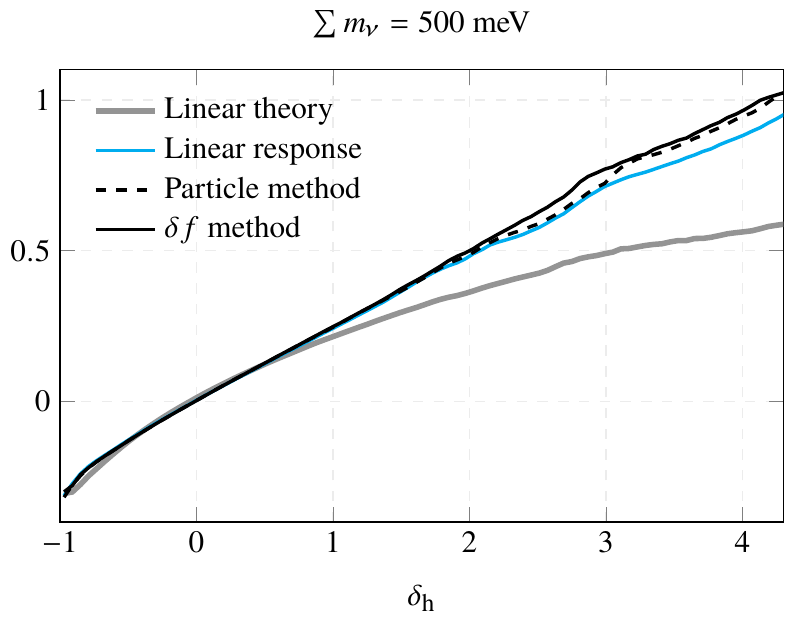}
	}
	\caption{Neutrino bias relative to dark matter halos with virial mass, $M_\text{cdm+b}>10^{12}M_\odot$, on scales, $R=30h^{-1}$ Mpc, computed with four different methods from the Gpc simulations at $z=0$. The $\bar{\delta}_\nu(\delta_\text{h})$ relationship is approximately linear with slope equal to the neutrino bias $b$.}
	\label{fig:nubias_relation}
\end{figure*}

\subsection{Neutrino bias}

On larger scales, the neutrino density field can be reconstructed from the density of halos for a given neutrino mass spectrum \citep{inman15}. We therefore construct the halo overdensity field, 
\begin{align*}
	\delta_\text{h}(x) = \frac{n_\text{h}(x)-\bar{n}_\text{h}}{\bar{n}_\text{h}},
\end{align*}

\noindent
by calculating the number density, $n_\text{h}(x)$, of halos and the mean density, $\bar{n}_\text{h}$, at $z=0$ in our Gpc simulations identified using the halo finder \textsc{velociraptor} \citep{elahi19}. We restrict attention to halos with virial mass,  $M_\text{cdm+b}>10^{12}M_\odot$, and smooth $\delta_\text{h}$ and $\delta_\nu$ with a tophat filter of comoving radius $R=30h^{-1}$ Mpc. Following \citet{yoshikawa20}, we study the mean neutrino density at constant halo density $\bar{\delta}_\nu(\delta_\text{h})$, defined in terms of the joint probability density function $P(\delta_\nu,\delta_\text{h})$ as
\begin{align*}
\bar{\delta}_\nu(\delta_\text{h}) = \int\mathrm{d}\delta_\nu\delta_\nu P(\delta_\nu,\delta_\text{h}).
\end{align*}

\noindent
This relationship is close to linear with slope equal to the neutrino bias, given by
\begin{align*}
	b = \frac{\langle \delta_\nu\delta_\text{h}\rangle}{\langle\delta_\text{h}^2\rangle}.
\end{align*}

\noindent
The degree of nonlinearity is captured by
\begin{align*}
    \epsilon_\text{nl}^2 = \frac{\langle\delta_\text{h}^2\rangle\langle\bar{\delta}_\nu^2\rangle}{\langle\bar{\delta}_\nu\delta_\text{h}\rangle^2}-1,
\end{align*}

\noindent
which satisfies $\epsilon_\text{nl}=0$ if and only if the slope of $\bar{\delta}_\nu(\delta_\text{h})$ is independent of $\delta_\text{h}$. The scatter around the biasing relationship is characterized by the stochasticity,
\begin{align*}
    \epsilon_\text{stoch}^2 = \frac{\langle\delta_\text{h}^2\rangle\langle(\delta_\nu - \bar{\delta}_\nu)^2\rangle}{\langle\bar{\delta}_\nu\delta_\text{h}\rangle^2}.
\end{align*}

\noindent
The nonlinearity and stochasticity are related to the correlation coefficient,
\begin{align*}
    r_{\nu,\text{h}} = \frac{\langle\delta_\nu\delta_\text{h}\rangle}{\sqrt{\langle\delta_\nu^2\rangle\langle\delta_\text{h}^2\rangle}},
\end{align*}

\noindent
via $r_{\nu,\text{h}}\simeq(1+\epsilon_\text{nl}^2+\epsilon_\text{stoch}^2)^{-1/2}$. This model is analogous to the nonlinear stochastic galaxy biasing model of \citet{taruya00,yoshikawa01}. We compute the four quantities $(b,\epsilon_\text{nl}^2,\epsilon_\text{stoch}^2,r_{\nu,\text{h}})$ for each of the methods under consideration. The results are listed in Table~\ref{tab:bias} and the biasing relationship is shown in Fig.~\ref{fig:nubias_relation}. As expected on these large scales, we find good agreement with differences of a few percent in the bias. The greater the level of neutrino clustering resolved by a given method, the greater the bias $b$ and correlation $r_{\nu,\text{h}}$. The stochasticity follows the opposite pattern. The nonlinearity follows no such pattern, but is very small in each case except (amusingly) for the linear theory runs. This is because linear theory does not resolve neutrino halos, causing the $\bar{\delta}_\nu(\delta_\text{h})$ relation to level off in the high density tail.

The bias $b=0.103$ for the 100 meV scenario is in excellent agreement with the bias $b=0.071$ found by \citet{yoshikawa20}, when the difference in mass ordering is factored in using the approximately linear relationship between neutrino mass and bias in their results. \citet{yoshikawa20} do not consider neutrino masses beyond 400 meV, but our finding of $b=0.256$ for 500 meV is slightly lower than expected when extrapolating from their results. We also find a larger stochasticity and smaller correlation than  might be expected, although the small nonlinearities agree. Given the mutual agreement between the different runs in Table~\ref{tab:bias}, these differences are unlikely to be due to our choice of neutrino method. Differences in the $N$-body code or the identification of halos could also affect this comparison.

\begin{table}
	\centering
	\caption{Neutrino bias relative to dark matter halos on scales $R=30h^{-1}$ Mpc. Listed are the bias, $b$; nonlinearity, $\epsilon_\text{nl}^2$; stochasticity, $\epsilon_\text{stoch}^2$; and the correlation coefficient, $r_{\nu,\text{h}}$.}
    \begin{tabular}{|c|l|c|c|c|c|}
        \hline
        \hline
        & \multicolumn{1}{c|}{Method} & \multicolumn{1}{c|}{$b$} & \multicolumn{1}{c|}{$\epsilon_\text{nl}^2$} & \multicolumn{1}{c|}{$\epsilon_\text{stoch}^2$} & \multicolumn{1}{c|}{$r_{\nu,\text{h}}$}\\
        \hline
        \parbox[t]{2mm}{\multirow{4}{*}{\rotatebox[origin=c]{90}{$100$ meV}}} & $\delta f$ method & $0.1032$ & $0.0022$ & $0.4883$ & $0.8195$\\
        & Particle method & $0.1028$ & $0.0021$ & $0.4955$ & $0.8176$\\
        & Linear response & $0.1015$ & $0.0022$ & $0.5065$ & $0.8146$\\
        & Linear theory & $0.0987$ & $0.0206$ & $0.5878$ & $0.7889$\\
        \hline
        \parbox[t]{2mm}{\multirow{4}{*}{\rotatebox[origin=c]{90}{$500$ meV}}} & $\delta f$ method & $0.2556$ & $0.0014$ & $0.1969$ & $0.9137$\\
        & Particle method & $0.2546$ & $0.0017$ & $0.1927$ & $0.9152$\\
        & Linear response & $0.2502$ & $0.0019$ & $0.2031$ & $0.9112$\\
        & Linear theory & $0.2404$ & $0.0257$ & $0.2902$ & $0.8719$\\
        \hline
        \hline
    \end{tabular}
    \label{tab:bias}
\end{table}

\begin{figure*}
	\normalsize
	\subfloat{
        \includegraphics{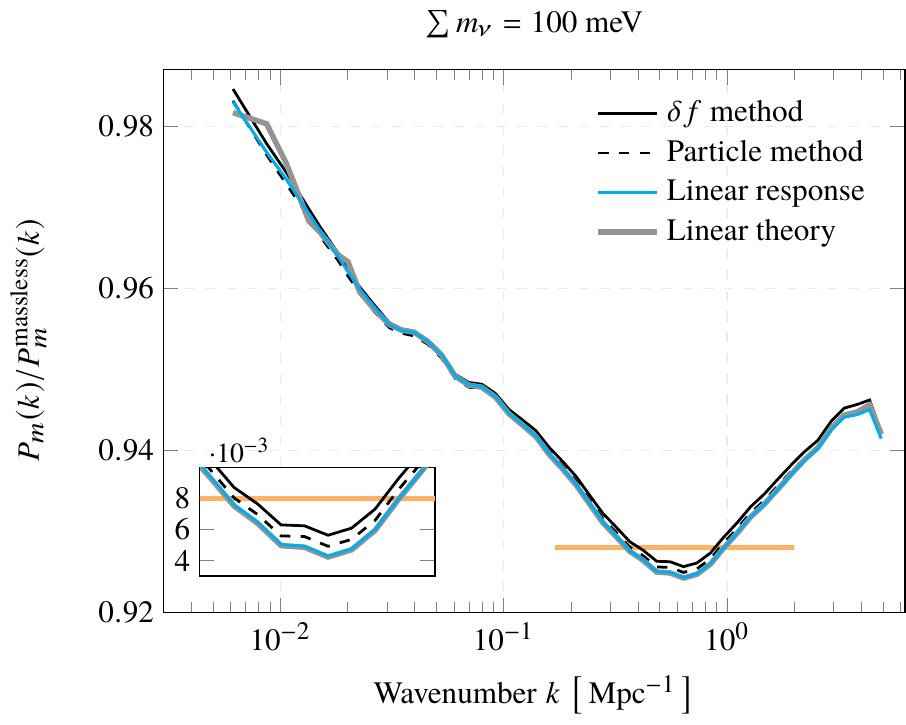}
	}
	\subfloat{
        \includegraphics{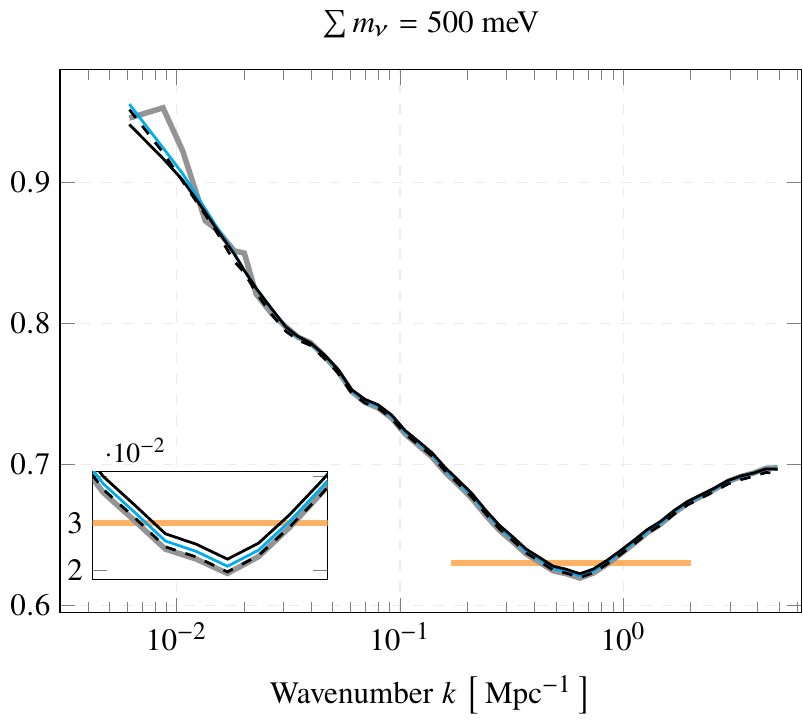}
	}
	\caption{Total matter power spectra at $z=0$, relative to a massless neutrino cosmology. The plots are based on $(1024\text{ Mpc})^3$ simulations with $N_\text{cb}=1024^3$ and (for the particle and $\delta f$ methods) $N_\nu=1024^3$ simulation particles. The horizontal line is the empirical fitting formula, $\Delta P/P = -9.8f_\nu$.}
	\label{fig:mpower}
\end{figure*}

\subsection{Matter power spectrum}

The suppression of the total matter power spectrum at $z=0$, relative to a massless neutrino cosmology, is shown in Fig.~\ref{fig:mpower}. We see that all methods are in excellent agreement and reproduce the famous spoon-like feature, which has recently been explained in terms of the halo model \citep{hannestad20}. The linear theory method disagrees on the largest scales, because its neutrino component is not affected by cosmic variance unlike the other methods. The differences between the methods are otherwise most pronounced around $k=0.6\;\text{Mpc}^{-1}$, where the suppression is largest. The inset graphs zoom in on these scales. For both neutrino masses, the $\delta f$ method predicts a smaller suppression than the particle and linear methods. This is in line with expectation, as the additional small-scale neutrino clustering slightly offsets the suppression. Similarly, the pure linear theory method predicts the least neutrino clustering and the largest suppression. It is interesting to see that the particle and $\delta f$ methods do not agree for $\Sigma m_\nu=500$ meV, despite having similar neutrino power spectra at $z=0$. This is most likely due to shot noise at high redshift, highlighting the importance of using a hybrid type method. The differences between the methods are at the permille level, corresponding to a shift in neutrino mass of several meV. In absolute terms, the differences are larger for $\sum m_\nu=500$ meV, but less important overall.

The horizontal line corresponds to the empirical fitting formula, $\Delta P/P=-9.8f_\nu$ \citep{brandbyge08}. Compared to this formula, we find a slightly greater suppression in each case, regardless of the method used to model the neutrinos. For the $100$ meV simulations, this can be attributed to our use of the inverted mass ordering. The $\sum m_\nu=500$ meV case is perhaps more surprising, but seems to be in line with recent works. For example, \citet{partmann20} find increasingly larger differences with the fitting formula for increasing masses, although they do not consider models with $\sum m_\nu>300$ meV.

Globally, the agreement between these very different methods is an encouraging sign and suggests that we have a good handle on the effects of massive neutrinos on the matter power spectrum. The differences, at most a few permille, may perhaps be relevant when trying to distinguish the effects of individual neutrino masses \citep{wagner12}.

\section{Higher order \lowercase{$\delta f$} methods}\label{sec:higher}

The performance of the $\delta f$ method scales with the correlation between the nonlinear solution $f(x,p,t)$ and the background model $f_0(x,p,t)$, so it is worth investigating other background models. We can go beyond the $0^\text{th}$ order Fermi-Dirac model by including the linear theory prediction. In that case, the distribution function can be written as
\begin{align}
	f_0(x,p,t) = f_\text{FD}(x,p,t) \left[1 + \Psi(x,p,t)\right], \label{eq:higher_f}
\end{align}

\noindent
where the perturbation is decomposed into multipole moments \citep{ma95},
\begin{align*}
	\Psi(k,p,t) = \sum_{\ell=0}^\infty (-1)^\ell (2\ell+1)\Psi_\ell(t,k,q)P_\ell(\hat{k}\cdot\hat{n}).
\end{align*}

\noindent
Here, $P_\ell(\cdot)$ are Legendre polynomials and the coefficients $\Psi_\ell$ satisfy an infinite hierarchy of moment equations. The Legendre representation yields simple expressions for the first few fluid moments, but is cumbersome for evaluating the distribution function itself. For our purposes, it is more convenient to use the following monomial representation
\begin{align*}
    \Psi(\mathbf{k},\hat{n},q,\tau) &= \sum_{\ell=0}^\infty i^\ell \Phi_\ell(\mathbf{k},q,\tau)(\mathbf{k}\cdot\hat{n})^\ell,
\end{align*}

\noindent
where for a given $\ell_\text{max}$, the odd (even) $\Phi_\ell(x,q,t)$ can be expressed in terms of all the odd (even) $\Psi_m(x,q,t)$ with $m\leq \ell$. See Appendix~\ref{sec:monomial} for details. With this choice of background model, the density integral becomes
\begin{align*}
	\rho(x) &= \bar{\rho}\left[1+\delta_\nu(x)\right] + \sum_{i=1}^N \sqrt{m^2+p_i^2}\,w_i\,\delta^{(3)}(x-x_i),
\end{align*}

\noindent
with particles weights $w_i = \delta f/f$ and $f_0$ given by \eqref{eq:higher_f}. Here, $\delta_\nu(x)$ is the linear neutrino overdensity, which is calculated using \textsc{class}. The effect of the $\delta_\nu$ perturbation should now be included in the long-range force calculation.

\begin{figure*}
	\normalsize
	\centering
	\includegraphics{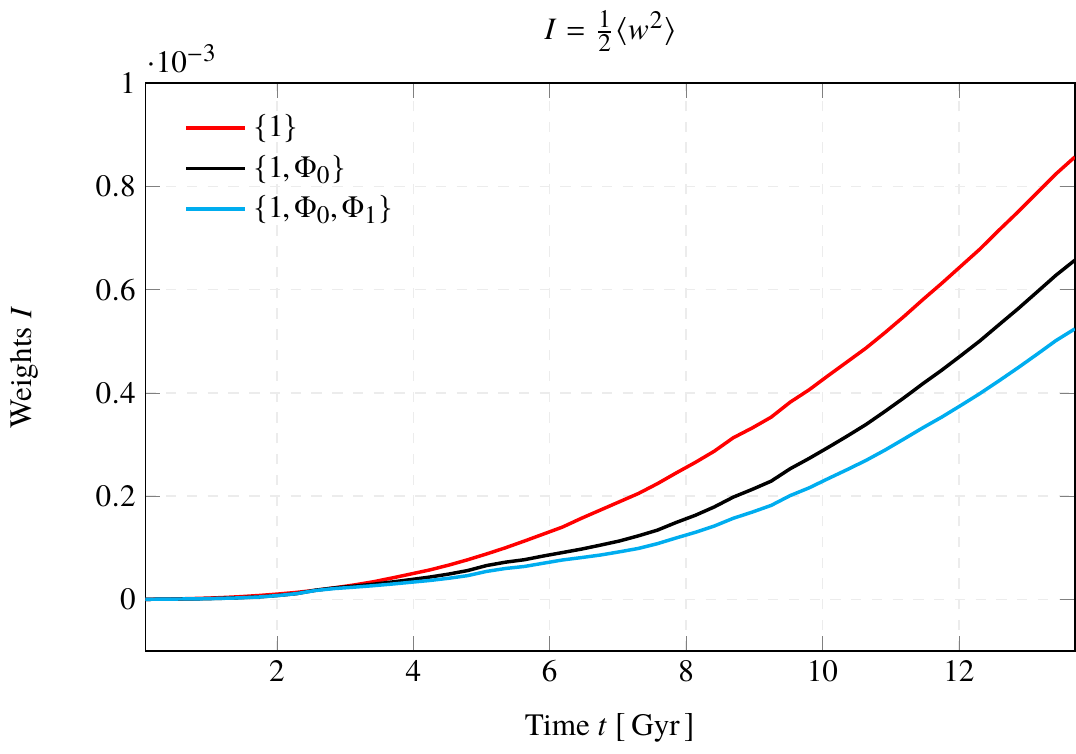}
	\caption{Evolution of the weights, or the effective reduction in shot noise, parametrized by the $I$-factor $I=\tfrac{1}{2}\langle w^2\rangle$, when including higher order perturbations: $\Phi_0$ (density), and $\Phi_1$ (energy flux).}
	\label{fig:weight_evol_higher}
\end{figure*}

As shown in Fig.~\ref{fig:weight_evol_higher}, adding the multipoles $\Phi_0$ and $\Phi_1$ significantly improves the correlation and therefore reduces the shot noise by almost 50\%. It is likely that higher order terms could contribute meaningfully too, as the multipole expansion converges only slowly. However, most of the gain is due to the $0^\text{th}$ order term, which on its own is much easier to implement.

\section{Discussion and conclusions}\label{sec:fin}

Shot noise in $N$-body simulations is a major obstacle to modelling the nonlinear evolution of light relic neutrinos. In this paper, we demonstrate that the $\delta f$ method, which decomposes the neutrino distribution into an analytically tractable background component,  $f_0$, and a nonlinear perturbation, $\delta f$, carried by the simulation particles, is an effective variance reduction technique. The reduction in shot noise scales with the dynamic particle weights, parametrized by $I=\tfrac{1}{2}\left\langle w^2\right\rangle$. Because the weights are negligible until very late times, the simulation is mostly immunized against the effects of shot noise. Furthermore, shot noise is greatly reduced even at $z=0$, which makes it possible to resolve neutrino clustering down to much smaller scales than is possible with conventional methods. Using higher order versions of the $\delta f$ method, which incorporate additional information from perturbation theory, shot noise can be reduced by another factor of $\mathcal{O}(2)$, and possibly more if moments $\ell>2$ are included. Additional reduction in shot noise is possible by carefully tuning the sampling distribution of the marker particles.

The reduction in shot noise is more significant for smaller neutrino masses, because faster particles deviate less from their initial trajectory, resulting in smaller weights. This is fortunate as shot noise is most problematic for the fastest neutrinos. More generally, particles whose trajectories are not perturbed have negligible weights, whereas particles that are captured by halos have appreciable weights. This is again fortunate, because particles are needed in the vicinity of halos where grid methods tend to fail, while the unperturbed particles contain no information and contribute only noise. In between these extremes, particles will have intermediate weights. In this way, the $\delta f$ method ensures an optimal combination of particles and background.

The method can in principle be combined with any grid or fluid background model to obtain an optimal hybrid method. Any simulation that evolves neutrino particles can be extended with a weighting step to minimize the shot noise as outlined in section \ref{sec:implement}. It is not necessary, as was done here, to evolve the neutrino particles from the beginning. The $I$-statistic from a reference simulation can be used to gauge when the neutrinos become nonlinear and at what point they can safely be introduced (see Fig.~\ref{fig:weights_time}).

We know from neutrino oscillations that at least one neutrino has a mass $M_\nu\gtrsim 0.05$ eV. Our results indicate that even for masses close to that bound, neutrinos are not particularly well modelled by linear approximations. For instance, the linear response neutrino power spectrum is off by 10\% (60\%) at $k=0.1\;\text{Mpc}^{-1}$ $(k=1\;\text{Mpc}^{-1})$ at $z=0$, and the pure linear theory prediction is off by 14\% (96\%). Because the neutrinos make up only a small fraction of the total mass, the effect on the matter power spectrum is at most a few permille. This is the level at which the mass splittings are important \citep{wagner12}. Other statistics may be affected at a greater level, particularly if they are more sensitive to neutrino effects. For example, we have shown that the neutrino bias relative to dark matter halos is affected at the percent level on $30h^{-1}$ Mpc scales. In addition, some novel probes may require accurate modelling of the neutrino dynamics around halos, such as the neutrino-induced dynamical friction \citep{okoli17} and torque \citep{yu19} on halos. By reducing shot noise without neglecting nonlinear terms, the $\delta f$ method makes it feasible to calculate these effects even for the lightest neutrinos.

\section*{Acknowledgements}

We acknowledge support from the European Research Council through ERC Advanced Investigator grant, DMIDAS [GA 786910] to CSF. WE is supported by the ICC PhD Scholarship. SP acknowledges partial support from the European Research Council under ERC Grant NuMass (FP7-IDEAS-ERC ERC-CG 617143), the European Unions Horizon 2020 research and innovation programme under the Marie Sklodowska-Curie grant agreement No. 690575 (RISE InvisiblesPlus) and No. 674896 (ITN Elusives). BL is supported by the European Research Council (ERC) through ERC starting Grant No.~716532, and STFC Consolidated Grant (Nos.~ST/I00162X/1, ST/P000541/1). This work was also supported by STFC Consolidated Grants for Astronomy at Durham ST/P000541/1 and ST/T000244/1. This work used the DiRAC Data Centric system at Durham University, operated by the Institute for Computational Cosmology on behalf of the STFC DiRAC HPC Facility (www.dirac.ac.uk). This equipment was funded by BIS National E-infrastructure capital grants ST/K00042X/1, ST/P002293/1, ST/R002371/1 and ST/S002502/1, Durham University and STFC operations grant ST/R000832/1. DiRAC is part of the National e-Infrastructure.

\bibliographystyle{mnras}
\bibliography{main}

\appendix

\section{Elliptical sine wave solution}\label{sec:sinedetails}

We define an integral of motion
\begin{align*}
	E(x,p,t) = \tfrac{1}{2}p^2 + \sin^2(x/2),
\end{align*}

\noindent
which is interpreted as the energy of the particle. Hence,
\begin{align*}
	p = \pm\sqrt{2E - 2\sin^2(x/2)}.
\end{align*}

\noindent
We have reduced the characteristic equations to
\begin{align*}
	\frac{\mathrm{d}x}{\mathrm{d}t} = \pm\sqrt{2E - 2\sin^2(x/2)}.
\end{align*}

\noindent
This equation is separable,
\begin{align*}
	\int\frac{\mathrm{d}x}{\sqrt{2E - 2\sin^2(x/2)}} &= \pm\int\mathrm{d}t.
\end{align*}

\noindent
Let $\tau$ be the time when $x(\tau)=0$. Putting in the integration limits, setting $u=x/2$, and factoring out $2E$, we obtain
\begin{align*}
	\frac{2}{\sqrt{2E}}\int_0^u\frac{\mathrm{d}u'}{\sqrt{1 - \sin^2(u')/E}} &= \pm\int_{\tau}^t\mathrm{d}t'.
\end{align*}

\noindent
The integral on the left defines the elliptic sine function
\begin{align}
	\mathrm{sn}\left(\pm\sqrt{E/2}(t - \tau)\right) = \sin u.\label{eq:sine_eom_sol}
\end{align}

\noindent
The elliptic sine satisfies the following trigonometric identities \citep{weisstein02}
\begin{align*}
	\mathrm{sn}(\theta)^2 + \mathrm{cn}(\theta)^2 = 1 \;\,\;\,\;\,\;\text{and}\;\,\;\,\;\,\; \frac{\mathrm{d}}{\mathrm{d}\theta}\mathrm{sn}(\theta) = \mathrm{cn}(\theta)\mathrm{dn}(\theta),
\end{align*}

\noindent
where $\mathrm{cn}(\theta)$ and $\mathrm{dn}(\theta)$ are the elliptic cosine and delta amplitude functions. Using these identities, one can confirm the solution \eqref{eq:sine_eom_sol}. To find the phase-space distribution at time $t$, we use the fact that $f(x,p,t)$ is constant along its characteristic curves. Hence, using the initial Gaussian distribution \eqref{eq:sine_ini} at time $t_0=0$, we find
\begin{align*}
	f(x,p,t) &= f(x_0,p_0,0)\\
			&= \frac{\rho_0}{\sqrt{2\pi\sigma^2}}\exp\left(-\frac{(p_0(x,p,t)-\bar{p})^2}{2\sigma^2}\right).
\end{align*}

\noindent
We need to express $p_0$ in terms of $x$, $p$, and $t$. First, we use conservation of energy to note that
\begin{align*}
	p_0^2 = p^2+2\sin^2(x/2)-2\sin^2(x_0/2).
\end{align*}

\noindent
What remains to show is how to express $\sin^2(x_0/2)$ in terms of $x$, $p$, and $t$. But this is simply, 
\begin{align*}
	\sin^2(x_0/2) &= \mathrm{sn}^2\left(\mp\sqrt{\tfrac{1}{4}p^2+\tfrac{1}{2}\sin^2(x/2)}\tau\right),
\end{align*}

\noindent
where the time, $\tau$, is given by
\begin{align*}
	\tau &= \mp\sqrt{2/E}\;\mathrm{arcsn}\left(\sin(x/2)\right) + t.
\end{align*}

\noindent
Here, we used the inverse of the elliptic sine function, $\text{arcsn}(x)=\theta$, with $x=\mathrm{sn}(\theta)$. It follows that
\begin{align*}
	\sin^2(x_0/2) &= \mathrm{sn}^2\left(\mathrm{arcsn}\left(\sin(x/2)\right) \mp\sqrt{\tfrac{1}{4}p^2+\tfrac{1}{2}\sin^2(x/2)}t\right).
\end{align*}

\noindent
Therefore, the distribution function is
\begin{align*}
f(x,p,t) &= \frac{\rho_0}{\sqrt{2\pi\sigma^2}}\exp\left(-\frac{1}{2\sigma^2}\left[h(x,p,t) - 2\bar{p}\sqrt{h(x,p,t)}+\bar{p}^2\right]\right),\\
h(x,p,t) &= k(x,p,1) - m(x,p,t)\\
k(x,p,t) &= p^2 + 2\sin^2(x/2)t\\
m(x,p,t) &= 2\mathrm{sn}^2\left(\mathrm{arcsn}\left(\sin(x/2)\right) \mp\sqrt{\tfrac{1}{4}k(x,p,t)}\right).
\end{align*}
    
\vspace{-2em}
\noindent
To find the density profile, $\rho(x,t)$, we integrate
\begin{align*}
	\rho(x,t) &= \int_{-\infty}^\infty f(x,p,t)\mathrm{d}p,
\end{align*}

\noindent
which can be done numerically. This gives the solution curves in Fig.~\ref{fig:sinedens}.

\section{Accurate calculation of neutrino moments}\label{sec:firebolt}

For completeness, we outline the linear theory calculation of the neutrino distribution function following \citet{ma95}. The neutrino phase-space distribution function can be written as
\begin{align}
    f(\mathbf{x},\mathbf{q},\tau) = f_0(q)\left[1+\Psi(\mathbf{x},q,\hat{n},\tau)\right],\label{eq:def1}
\end{align}

\noindent
where $f_0(q)$ is the homogeneous Fermi-Dirac distribution and the momentum $\mathbf{q}$ has been decomposed into a magnitude $q$ and a unit vector $\hat{n}$. We will express the momentum $q$ and energy $\epsilon=\sqrt{q^2+a^2m^2}$ in units of $k_b T$. In synchronous gauge, the perturbation $\Psi$ evolves as
\begin{align*}
    \dot{\Psi} + i\frac{qc}{\epsilon}(\mathbf{k}\cdot\hat{n})\Psi + \frac{1}{c^2}\frac{\mathrm{d}\ln f_0}{\mathrm{d}\ln q}\left[\dot{\eta} - \frac{\dot{h}+6\dot{\eta}}{2}(\hat{k}\cdot\hat{n})^2\right] = 0,
\end{align*}

\noindent
where we switched to momentum space and where overdots denote conformal time derivatives. Here, $h$ and $\eta$ are the scalar metric perturbations in synchronous gauge. To solve this equation, $\Psi$ is decomposed into a Legendre series
\begin{align}
    \Psi(\mathbf{k},\hat{n},q,\tau) = \sum_{\ell=0}^\infty (-i)^\ell (2\ell+1)\Psi_\ell(\mathbf{k},q,\tau)P_\ell(\hat{k}\cdot\hat{n}). \label{eq:def2}
\end{align}

\noindent
The Boltzmann equation then becomes an infinite tower of equations:
\begin{align}
    \dot{\Psi}_0 &= -\frac{qkc}{\epsilon}\Psi_1 + \frac{\dot{h}}{6}\frac{1}{c^2}\frac{\mathrm{d}\ln f_0}{\mathrm{d}\ln q}, \label{eq:b0}\\
    \dot{\Psi}_1 &= \frac{qkc}{3\epsilon}(\Psi_0 - 2\Psi_2), \label{eq:b1}\\
    \dot{\Psi}_2 &= \frac{qkc}{5\epsilon}(2\Psi_1 - 3\Psi_3) - \left(\frac{\dot{h}}{15}+\frac{2\dot{\eta}}{5}\right)\frac{1}{c^2}\frac{\mathrm{d}\ln f_0}{\mathrm{d}\ln q}, \label{eq:b2}\\
    \dot{\Psi}_\ell &= \frac{qkc}{(2\ell+1)\epsilon}\left[\ell\Psi_{\ell-1} - (\ell+1)\Psi_{\ell+1}\right], \;\,\;\,\;\,\;\,\text{for}\;\,\;\,\ell\geq3. \label{eq:b3}
\end{align}

\noindent
The hierarchy can be truncated at $\ell=\ell_\text{max}+1$ using the ansatz
\begin{align*}
    \Psi_{\ell_\text{max}+1} = \frac{(2\ell_\text{max}+1)\epsilon}{qkc\tau}\Psi_{\ell_\text{max}} - \Psi_{\ell_\text{max}-1}.
\end{align*}

The precision of this calculation is set by two parameters: the maximum multipole $\ell_\text{max}$ and the number of momentum bins $N_q$. By default, \textsc{class} partly relies on a set of fluid equations and partly on integrating the hierarchy, using $\ell_\text{max}=50$ and $N_q=28$ at the pre-set high precision settings \citep{lesgourgues11b}. The differences in the CMB anisotropies are at the permille level. However, the neutrino transfer functions have still not converged. To obtain converged results, \cite{dakin19} ran calculations with $N_q=2000$ bins and $\ell_\text{max}=2000$, which each required hundreds of CPU hours. This is to be contrasted with a default \textsc{class} run, which completes in seconds. To circumvent this computational cost, we use a different approach, which involves a post-processing step of \textsc{class} tables.

To quickly integrate the Boltzmann hierarchy for high $N_q$ and $\ell_\text{max}$, we note that the source terms in the evolution equations depend on the matter content only through the scalar potential derivatives $\dot{h}$ and $\dot{\eta}$, which can be calculated accurately with much lower settings\footnote{In the reference model with $N_q=28$ and $\ell_\text{max}=50$, relative errors in $\dot{h}$ and $(\dot{h}/3+2\dot{\eta})$ are of order $10^{-4}$. Although $\dot{\eta}$ still fluctuates at the several per cent level, this term is much smaller than $\dot{h}$.}. Therefore, we make the assumption that we can decouple the potential terms from most of the neutrino moments $\Psi_\ell$. We first evolve all source functions in \textsc{class} at a reasonable precision setting. This gives the metric perturbations $\dot{h}(k,\tau)$ and $\dot{\eta}(k,\tau)$, which we then take as given and use to integrate the multipole moments $\Psi_\ell$ at high precision where they are needed.

\section{Monomial basis for the distribution function}\label{sec:monomial}

Boltzmann codes can solve for the functions $\Psi_\ell(k,q,\tau)$. But evaluating the distribution function, $f(\mathbf{x},\mathbf{q},\tau)$, requires substituting these back into the definitions \eqref{eq:def1} and \eqref{eq:def2}. This presents a challenge as the $\Psi_\ell$ are large discretely sampled arrays of amplitudes that need to be convolved with the random phases. It would be prohibitively expensive to do this repeatedly for each term in the Legendre expansion. We therefore adopt the following scheme. First, we use the following representation of the $\ell$th Legendre polynomial,
\begin{align*}
    P_\ell(x) = 2^\ell\sum_{n=0}^\ell x^n\binom{\ell}{n}\binom{\frac{n+\ell-1}{2}}{\ell},
\end{align*}

\noindent
where the last factor is a generalized binomial coefficient. This allows us to expand $\Psi$ and collect monomial terms in $\hat{k}\cdot\hat{n}$. We write
\begin{align*}
    \Psi(\mathbf{k},\hat{n},q,\tau) &= \sum_{\ell=0}^\infty i^\ell \Phi_\ell(\mathbf{k},q,\tau)(\mathbf{k}\cdot\hat{n})^\ell,
\end{align*}

\noindent
where the functions $\Phi_\ell$ are defined by
\begin{align*}
    \Phi_\ell(\mathbf{k},q,\tau) = \frac{1}{k^\ell}\sum_{n=0}^\infty (-2)^n\binom{n}{\ell}\binom{\frac{n+\ell-1}{2}}{n}(2n+1)\Psi_n(\mathbf{k},q,\tau).
\end{align*}

\noindent
Note that we factored out the magnitude of $\mathbf{k}=k\hat{k}$ and write the expansion in terms of $(\mathbf{k}\cdot\hat{n})^\ell$ and not $(\hat{k}\cdot\hat{n})^\ell$. This is to facilitate taking derivatives, as shown below. The Fourier transform of $\Psi$ is
\begin{align*}
    \Psi(\mathbf{x},\hat{n},q,\tau) = \int\frac{\mathrm{d}^3k}{(2\pi)^3}\Psi(\mathbf{k},\hat{n},q,\tau)e^{i\mathbf{x}\cdot\mathbf{k}},
\end{align*}

\noindent
and similarly for the $\Psi_\ell$ and $\Phi_\ell$. We write the directional derivative along the unit vector $\hat{n}$ as $\mathcal{D}_{\hat{n}}=n^i\partial_{x_i}$. In other words,
\begin{align*}
    \mathcal{F}\left\{\mathcal{D}_{\hat{n}} \Phi_\ell(\mathbf{x},\hat{n},q,\tau)\right\} \Longleftrightarrow i(\mathbf{k}\cdot\hat{n})\Phi_\ell(\mathbf{k},q,\tau).
\end{align*}

\noindent
Hence, we obtain
\begin{align*}
    i^\ell \Phi_\ell(\mathbf{k},q,\tau)(\mathbf{k}\cdot\hat{n})^\ell \Longleftrightarrow \mathcal{F}\left\{\mathcal{D}_{\hat{n}}^\ell \Phi_\ell(\mathbf{x},q,\tau)\right\}.
\end{align*}

\noindent
And so, the overall perturbation, $\Psi$, is
\begin{align*}
    \Psi(\mathbf{x},\hat{n},q,\tau) &= \sum_{\ell=0}^\infty \mathcal{D}_{\hat{n}}^\ell \Phi_\ell(\mathbf{x},q,\tau).
\end{align*}

\noindent
A convenient numerical scheme is to store the Fourier transformed grids $\Phi_\ell(\mathbf{x},q,\tau)$, in which case we can evaluate the distribution function efficiently by taking finite differences.

\label{lastpage}
\end{document}